# The Initiation of Shear Band Formation in Deformed Metallic Glasses from Soft Localized Domains

*Xinyi Wang[1], Hao Zhang[1] †, Jack F. Douglas[2] †*

[1] Department of Chemical and Materials Engineering, University of Alberta, Edmonton, Alberta, Canada, T6G 1H9

[2] Material Measurement Laboratory, Materials Science and Engineering Division, National Institute of Standards and Technology, Gaithersburg, Maryland, USA, 20899

**Abstract**

It has long been thought that shear band (SB) formation in amorphous solids initiates from relatively 'soft' regions in the material in which large-scale non-affine deformations become localized. The test of this hypothesis requires an effective means of identifying 'soft' regions and their evolution as the material is deformed to varying degrees, where the metric of 'softness' must also account for the effect of temperature on local material stiffness. We show that the mean square atomic displacement on a caging timescale $<u^2>$, the 'Debye-Waller factor', provides a useful method for estimating the shear modulus of the entire material and, by extension, the material stiffness at an atomic scale. Based on this 'softness' metrology, we observe that SB formation indeed occurs through the strain-induced formation of localized soft regions in our deformed metallic glass free-standing films. Unexpectedly, the critical strain condition for SB formation occurs when the softness ($<u^2>$) distribution within the emerging soft regions approaches that of the interfacial region in its undeformed state, initiating an instability with similarities to the transition to turbulence. Correspondingly, no SBs arise when the material is so thin that the entire material can be approximately described as being 'interfacial' in nature. We also quantify relaxation in the glass and the nature and origin of highly non-Gaussian particle displacements in the dynamically heterogeneous SB regions at times longer than the caging time.

*Corresponding authors: hao.zhang@ualberta.ca; jack.douglas@nist.gov

## Introduction

Metallic glasses (MG) have been intensively researched as a promising new class of materials for many applications, but the inherent brittleness of these materials, and other technical problems have limited their development [1–4]. In particular, MGs tend to readily undergo catastrophic fracture following the formation of highly localized shear bands (SB) in the material in which deformation becomes highly localized, precipitating material failure [1–3]. Given the practical importance of SBs in MGs, the process and mechanisms of SB formation have been extensively studied by both experimental and computational methods [5–8]. Experimental methods of studying SBs normally cannot resolve many aspects of SB formation and evolution because the nanoscale width and rapid evolution of these structures make measurements extremely challenging, and the long computational timescales required, and the inherently low temperatures characteristic of MG materials make the problem of studying SB formation and evolution also extremely difficult from a simulation standpoint. Since the timescale of the process is extremely short, and the length-scale is initially on the order of nm, we try to gain insights in SB formation and evolution through molecular dynamics (MD) simulation since this method at least has the requisite time and spatial resolution for studying this phenomenon.

Shear bands form spontaneously in MG materials at temperatures well below their glass transition temperature $T_g$ where molecular diffusion is unmeasurably slow by simulation, yet the formation of shear bands hints at the emergence of appreciable mobility in the material under deformation. The nature of 'mobility', and how to quantify it, under conditions where one cannot normally estimate the rate of molecular diffusion are clearly basic issues that need to be addressed. There is also a universal geometrical character to the SBs themselves that invites theoretical explanation. SBs in diverse materials take the form of ribbon-like thin regions that arise from the emergence of localized layer-like regions of the material in which relatively large strains arise, hence the term 'shear band'. This is the primary deformation mode of metallic glasses [9] that is of recognized critical importance for the mechanical properties of these materials in relation to their performance in applications. There is a general view in the MG community that if the mechanism of shear localization could be understood, along with the subsequent formation of mature SBs, then this knowledge might be used to engineer tougher MG materials that are less prone to this elastic instability. It is this ultimate goal of developing design rules for making metallic glasses for many applications that animate our work. It is notable that many aspects of SB formation in glass materials are remarkably insensitive to the chemical nature of the material. For example, the observations of Argon et al. [10] on shear banding in polystyrene in its glass state are remarkably similar to those for metallic glasses and the similarity between shear banding in polymeric and metallic glass materials has also been emphasized by Shavit and Riggleman [11]. This commonality in the phenomenology of shear band formation in such different materials gives us hope that there are general principles to be discovered about SB formation and yield in glassy materials that are invariant to the material substance.

Our own approach to this problem is heavily informed by ideas and computational tools derived from quantifying mobility fluctuations and collective motion arising in glass-forming (GF) liquids as these materials approach their glass transition from *above*. It is initially seemed reasonable to us to suppose that these dynamical structures, and atomic clustering associated with these mobility fluctuations, become kinetically trapped in the glass state. In the previous simulations, we and others have found [12,13] that the atoms of MGs and other model GF liquids tend to form string-like clusters of atoms corresponding to locally energetically preferred configurations and that the particles in these clusters tend to be relatively immobile in the fluid state under conditions in which the diffusion coefficients of different types of atoms in the material can be quantified by simulation. Ma [14] has recently provided a detailed review of locally preferred packing in metallic glass materials and Ding et al. [15] discuss this local packing geometry for the specific MG material studied in the present work. Many simulations have also shown that these relatively well-packed particles having an energetically preferred spatial configuration can be identified by a Voronoi analysis of the local environment of the atoms where it is found that the Voronoi cells are organized in highly correlated polymer-like structures in space. This is clearly a kind of topological 'ordering' process, as evidenced by the corresponding parallel drop in the fluid entropy in polymerizing systems and GF liquids [16], so that the material structure is not really 'random'. Of course, not all the atoms can participate in this locally preferred packing, and the atoms in the not so well-packed regions are energetically frustrated [14] in a fashion highly analogous to the 'amorphous regions' in polycrystalline materials [17]. We have found in our previous simulations that the dynamics of both metallic and polymeric GF liquids exhibit dynamic interpenetrating structures in which particle mobility is excessively high and low in these regions, respectively, and, in the liquid state the particles undergo a constant exchange between these dynamical states. We may then expect some remnants of this dynamical and structural heterogeneity of the liquid state to persist in the glass state below the glass transition temperature, $T_g$. Under deformation conditions, where mechanical rather than thermal energy should 'activate' particle motion, and we also expect active dynamic heterogeneity to re-emerge upon increasing stress until the material locally 'melts'. The present work pursues how 'dynamic heterogeneity' might influence SB formation in a model Zr-Cu metallic glass material. We naturally bring the tools that we previously brought to studying metallic GF liquids to study the dynamics of our simulated MG in its glass state. We note recent work has shown that deforming glass-forming materials leads to changes roughly equivalent to increasing $T$ [18] and another goal of our work is check this possibility.

The concept of dynamical heterogeneities has independently been introduced into phenomenological descriptions of the dynamics of MG and other glass materials. In particular, shear transformation zones (STZs) have been invoked to model the deformation processes of MG, these proposed structures being roughly the 'defect' counterpart of dislocations in deformed crystalline materials in their significance of the mechanical properties such as strain hardening, etc. The basic 'function' of the proposed STZs is to enable localized deformation in the material that allows macroscopically deformation through the action of numerous hypothetical regions of this kind spread throughout the material. Extensive research on metallic glasses has revealed that the relaxation process obtained from fitting models of the mechanics, with local properties assumed for STZs to mechanical measurements, [19] has indicated that the activation energy of the STZs is thermally activated where the thermal activation energy is highly correlated in general with the Johari-Goldstein (JG) relaxation process. It is widely appreciated that this relaxation process is the dominant relaxation process in mechanical and dielectric measurements of materials in their glass state [20]. This is highly encouraging from the standpoint of the present study because we have recently found that the JG relaxation process in model Al-Sm metallic glass materials can be quantitatively described by collective motions that have been observed in GF liquids [21,22]. There would then appear to be some prospect of obtaining a more fundamental understanding and precise specification of the rather abstract STZs. Recent works [23–25] have indicated that well-defined elastic heterogeneities in the stress field develop in deformed metallic glasses and we may expect the heterogeneities that we tentatively identify with the STZs to likewise exhibit stress multipole interactions. We hope to consider this problem in future work in simulations dedicated to elucidating this hypothesis.

Recently, Cao et al. quantified the network of icosahedral clustering mentioned and suggested its important role in the initiation of shear band formation [19]. This work reminds us that it is the organization of the locally well-packed particles in the material that is ultimately responsible for the 'strength' (shear modulus) of the material and the alteration of these structures under deformation can be expected to be an important contribution to the deformation dependence of material properties, in addition to alterations of the less well-packed regions exhibiting relatively high local deformability. Evidence of the importance of the deformation of these immobile particle clusters was emphasized in the magnitude of deformation-induced acceleration of structural relaxation in glass-forming liquids under steady shear conditions [18]. Both classes of 'dynamic heterogeneity', highly mobile and immobile particle clusters, can be expected to contribute to changes of materials properties under deformation.

One of the general trends that arises in liquids as they are cooled towards their glass-transition is a general increase in the scale of collective motion as the shear rigidity tends to increase [26] and this tendency at fixed $T$ can be expected to be contravened by material deformation. This tight interrelationship between cooperative motion and rigidity makes measures of rigidity particularly useful in the quantification of the dynamics of condensed materials. Since shear bands arise locally in the material, we are interested in local measures of rigidity that lend themselves to measurement and quantitative determination by simulation. The Debye-Waller factor (DWF), $<u^2>$ has been proposed as a measure of local material stiffness that has a highly predictive value in relation to understanding the material dynamics. This quantity is defined as the mean square displacement $<r^2>$ after a fixed 'caging' or decorrelation time $t_0$ characterizing the crossover from ballistic to caged atomic motion in the liquid dynamics [27]. $<u^2>$ is considered a 'fast dynamics' property since $t_0$ is typically on the order of a ps in molecular and atomic fluids. Remarkably, this dynamical property defined on this very short timescale has shown great value in estimating the structural relaxation time $\tau_\alpha$ on the timescale at the glass transition temperature $T_g$ where $\tau_\alpha$ is typically on the order of a min. This type of interrelationship has been explored in a range of materials and thus has some generality [18,28–32]. We note that previous studies have focussed on $<r^2>$ and various measures of local effective shear modulus in connection with the quantification of the onset of SB formation [33,34], but we are not aware of any previous study based on $<u^2>$.

In an early simulation study of $<u^2>$ in relation to relaxation in GF liquids, Starr et al. established a near proportionality between $<u^2>^{3/2}$ and 'rattle free volume' $<v_f>$, defined by the volume explored by the particle rattling about in a cage defined by the surrounding particles in the material [35,36]. Note that the inertial energy of the particles is a significant contributor to this type of free volume so that this quantity is quite different, even qualitatively in some cases, from estimates of 'free volume' based on material 'structure', e.g., estimates based on Voronoi volume, etc. We may expect $<u^2>$, which provides a useful local measure of stiffness, as a well as local mobility in terms of the mean amplitude of motion ('dynamic free volume'), to provide a useful metric for studying the emergence of local soft spots in deformed materials and the coalescence of soft regions into a shear band. Our previous studies of GF liquids have repeatedly shown that $<u^2>$ can inform on local mobility gradients that are not apparent structurally [37]. Previous work has shown that the magnitude of $<u^2>$ is often dominated by the motion of a relatively small fraction of particles in the system exhibiting string-like collective motion [38], and these dynamic structures are candidate structures for describing the STZs noted above. We may then expect to gain some important new insights in the formation and evolution of shear bands in deformed MGs by simply observing the evolution of the field of values of $<u^2>$. We find that this is indeed the case.

In the current paper, we examine the initiation and evolution of the formation of SB in $Cu_{64}Zr_{36}$ MG materials under uniaxial tensile loading with different sizes and temperatures. First, using commonly used local von Mises strain measure, we observed numerous areas with relatively large local strain spontaneously arises as the stress increases. Those areas grow and coalescence in particular band upon the stress approaching the critical value for the formation of SB, phenomenologically resembling early and late stage of spinodal decomposition. We then examined whether the common dynamic heterogeneity types found in the cooled GF liquid approaching the glass transition from above also exist in the glass state. Since the systems do not have pre-existing regions with stress concentration, we found that regions with relatively large $<u^2>$ first appeared simultaneously at many places within the material, but these regions evolved in their organization into a shear band become concentrated into a band-like domain. We find that when the SB has fully formed, almost all the particles inside these regions are 'interfacial-like' (i.e., atomic mobility is equivalent to the atomic mobility in the interfacial region at the undeformed state), which enables the relatively facile deformation of these regions. We then define a precise local measure of material stiffness to investigate the initiation of shear band formation. The local stiffness distribution in the SB region and the whole system both approach to the distribution of the interfacial region at the undeformed state. In particular, as the material strain increased, the fraction of mobile particles progressively grew at relatively random positions within the incipient shear band region. In addition, we found that the van Hove function describing atomic displacement in a highly deformed metallic glass exhibits rather typical behaviour for the type of elastically turbulent system observed during the formation of SB.

## Simulation Methodology

Molecular dynamics simulations were implemented to examine the formation of shear band in $Cu_{64}Zr_{36}$ metallic glass systems. The MD simulations were carried out using Large-scale Atomic/Molecular Massively Parallel Simulator (LAMMPS), developed by the Sandia National Laboratories [39]. The atomic interaction in metallic glasses systems was described by semi-empirical potentials optimized to reproduce the measured static structure factor and other equilibrium properties of Cu-Zr alloys [40]. It has been demonstrated in previous studies that this potential provided reliable descriptions of both structural and dynamic properties. In the present study, the representative alloys $Cu_{64}Zr_{36}$ were chosen based on extensive research and studies show that this alloy possesses good glass-forming ability near the eutectic points of the alloys.

Next, we describe the simulation methods for MD simulations of metallic glasses. First, we chose four different length scales of $Cu_{64}Zr_{36}$ alloys: $(180 \times 30 \times 30)$ Å ($\approx 10,125$ atoms), $(180 \times 90 \times 30)$ Å ($\approx 30\,375$ atoms), $300 \times 150 \times 30$ Å ($\approx 84\,375$ atoms), and $(600 \times 300 \times 60)$ Å ($\approx 675,000$ atoms) in X, Y, and Z directions. For the latter three simulation cells, the dimension in X and Y was chosen to keep the aspect ratio constant. Taking $(600 \times 300 \times 60)$ Å as an example, we begin with a perfect Cu single crystal containing 675 000 atoms. Then 36 % of the Cu atoms are randomly replaced by Zr atoms. The mixture was then heated from 300 K to 2000 K and kept at 2000 K for 5 ns to achieve a structural homogeneous glass-forming liquid. After that, the system was cooled down to 50 K in 19.5 ns, given a cooling rate of 100 K/ns. During this process, NPT (constant number of atoms, constant pressure, and constant temperature) ensemble was implemented with zero pressure and periodic boundary conditions. The constant pressure was controlled by the Parrinello-Rahman algorithm [41], and temperature was kept constant using Nose-Hoover thermostat method [42,43]. After obtaining the metallic glass sample, 30 Å vacuum spaces were added on both sides along the Y-axis.

The deformation of the sample was achieved by loading the sample under uniaxial tensile loading along X-axis using a constant strain rate of $1 \times 10^7$ / s to 15 %. Periodic boundary condition was applied on the X and Z-axis, while the free surface condition was applied on the Y-axis. The simulations were first carried out at 50 K to guarantee the formation of the shear band. To examine the temperature influence on the deformation, three more temperatures at 200 K, 300 K and 400 K were carried out on the system with $h$ = 300 Å. Same strain rate was applied with different temperatures. Atomic configurations were saved every 1 ps for future analysis.

**Results and Discussion**

**A.** *Uniaxial Tension Test on Cu$_{64}$Zr$_{36}$ Materials Having a Range of Thicknesses*

Figure 1 (a) shows stress-strain curves of Cu$_{64}$Zr$_{36}$ with four different thicknesses $h$ = 30 Å, 90 Å, 150 Å, and 300 Å. For $h$ = 90 Å, 150 Å, and 300 Å, stress overshot and dramatic drop have been observed after roughly strain at 7 %, which is known as a characteristic feature of localized deformation, similar to the findings in other literature [19,44,45]. After the stress drops, fluctuations in the stress-strain curves become noticeable. We find below that this phenomenon is related to an intermittent softening and stiffening of the shear band region. These soft areas have been identified as shear transformation zones in other literature [46]. Eventually, one dominant SB formed and led to a catastrophic failure in the sample.

The atomic shear strain $\eta^{\text{Mises}}$ was implemented to monitor the formation of the shear band, which was calculated using the equation[47]:

$$\eta^{Mises} = \sqrt{\varepsilon_{xy}^2 + \varepsilon_{xz}^2 + \varepsilon_{yz}^2 + \frac{(\varepsilon_{xx} - \varepsilon_{yy})^2 + (\varepsilon_{xx} - \varepsilon_{zz})^2 + (\varepsilon_{yy} - \varepsilon_{zz})^2}{6}} \qquad (1)$$

As shown in Figure 1 (b), atoms with the atomic shear strain larger than 0.2 have been marked as red to identify the deformation mode. Also, a non-localized deformation has been observed in the deformation map with $h$ = 30 Å, as well as in the stress-strain curves having no overshoot. This confirmed with the critical thickness $t_c$ = (3.33 ± 0.20) nm (note the variation was estimated based on multiple simulations) for $Cu_{64}Zr_{36}$ alloys proposed by Zhong et al. [44,45] that systems with smaller dimensions than the critical thickness exhibited a deformation transition mode from localized deformation to non-localized stress-induced uniform flow.

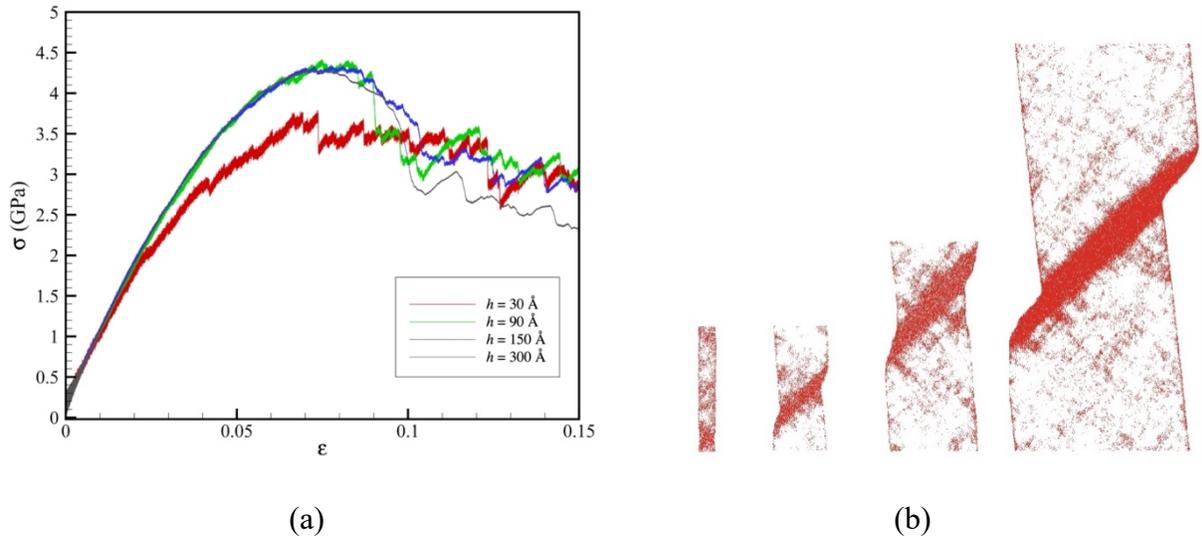

(a)                                                                (b)

**Figure 1.** Stress-strain curves (a) for $Cu_{64}Zr_{36}$ metallic glass materials under uniaxial extension at 50 K for films having a range of thicknesses and the onset of shear band formation. (b) Deformation maps of $Cu_{64}Zr_{36}$ with different thicknesses at $\varepsilon$ = 15 %. The red atoms are those with the von Mises local strain higher than 0.2. Corresponding from left to right: thickness $h$ = 30 Å, 90 Å, 150 Å, and 300 Å. The characteristic angle of the shear band near 45° corresponds to the direction in which the resolved shear stresses are maximum in this mode of deformation based on Schmid's law [48].

A series of snapshots of local von Mises strain directly demonstrating the formation of shear band is shown in Figure. 2. The sample was under uniaxial tensile loading without a stress concentrator. Von Mises local $\eta^{Mises}$ defined in eq. (1) were implemented to monitor the formation of shear band. The cut-off radius for von Mises strain calculation was chosen as the first minimum of the pair distribution function. As observed by Cao et al., systems without pre-existing stress concentrations tend to show multiple areas that undergo larger local deformation [19]. The shear band does not seem to initiate its growth from the surface, and then propagate along a 45° direction (maximum resolved shear stress direction) from this interface as we initially anticipated. (Below, we show that the dynamics of the interfacial region does have an impact on the onset of SB formation.) Instead, the SB gradually evidently builds up from a gradual accumulation of soft regions within the SB region. Once the band has fully formed in this way, it begins to widen and further soften, a pattern of growth that superficially resembles early and late stage of spinodal decomposition where local density or composition fluctuations first form spontaneously in local regions of the material, and this structure coarsens in time in a later coarsening stage. We see below, however, that the softness of the interior of the SBs closely matches the boundary of the material, and that the critical condition for SB formation is apparently related to a 'critical' condition at which this confluence of $<u^2>$ values occur. Thus, even if the SB is not 'nucleated' from the boundary the mechanical properties of the boundary and interfacial dynamics of the material appear to be highly relevant to the development of SBs.

Based on Griffith's theory, the propagation of shear band only succeeds when the elastic strain energy causing by the propagation is larger than the formation energy of the two free surfaces [49,50]. While the samples possess no pre-existing stress concentrator, fluctuations have been observed in terms of local strain. When the strain equals $\varepsilon = 8.0$ %, multiple areas with relatively large von Mises strain appeared within the sample. We believe that these areas are mainly responsible for mediating plastic deformation. When we increased the strain, these areas would soften or harden during the process. Eventually, one dominant shear band develops in which deformation becomes predominantly localized.

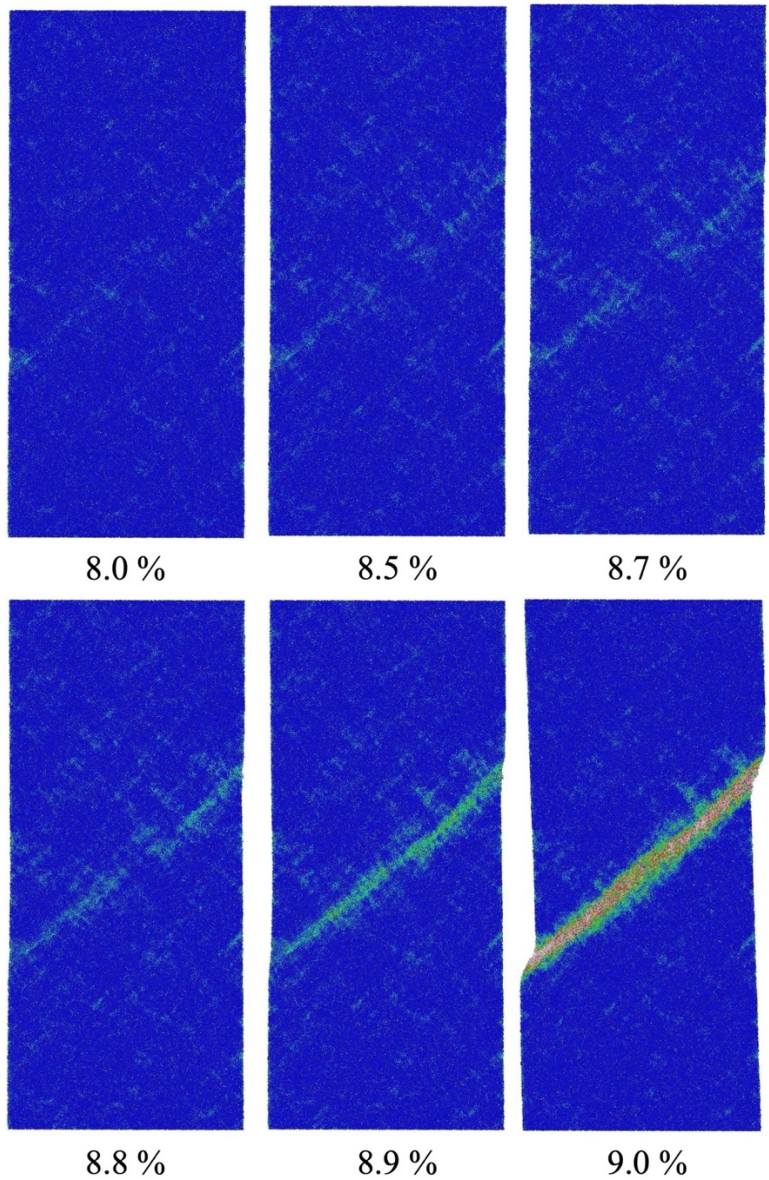

**Figure 2.** Snapshots of von Mises strain showing the formation of the shear band. Corresponding to strain $\varepsilon$ = 8.0 %, 8.5 %, 8.7 %, 8.8 %, 8.9 %, and 9.0 %.

## *B. Quantification of Local 'Mobility' and Relaxation in the Glass State*

At low temperatures in the metallic glass state, atomic diffusion is extremely slow and other mechanisms can be expected to mediate the structural relaxation process. The nature of these processes was not a priori clear to us when we started our investigation, so we first examined the dynamics of single particle displacement and collective atomic motion to determine what features are common to the cooled liquid regime approaching the glass transition from above or whether any entirely new modes of motion arise in the glass state. It is also seemed relevant to determine how the dynamics in the MG state depended on $T$ or whether the material properties strongly depended on the history of material preparation. The tentative answers to these questions guide how we analyze the onset and dynamics of shear banding when the material is subjected to deformation at low $T$.

We begin our analysis by considering the average mean square displacement (MSD) $<r^2(t)>$ of all atoms in our system as a function of time $t$ for different temperatures in Fig. 3(a). The MSD is defined as $\frac{1}{N}\sum_{i=1}^{N}\{[x_i(t) - x_i(0)]^2 + [y_i(t) - y_i(0)]^2 + [z_i(t) - z_i(0)]^2\}$ , where $(x_i(0), y_i(0), z_i(0))$ and $(x_i(t), y_i(t), z_i(t))$ are particle $i$'s initial and final positions after time $t$, respectively, and $N$ is the number of atoms. Notice the drift motion of all atoms due to the elastic deformation upon loading is subtracted. The $<r^2(t)>$ at a later time was calculated using the atomic position at $\varepsilon = 1$ % as a reference and calculated up to $\varepsilon = 2.9$ % (or equivalently 1.9 ns based on the current strain rate). It is ideal to use the configuration at $\varepsilon = 0$ % as the initial position for MSD calculation. Due to some unknown reason, the material exhibits stress exhibits fluctuations near 0 % strain. In order to avoid uncertainty in the calculation of MSD due to stress fluctuation, we instead choose the configuration at $\varepsilon = 1$%, where we believe it does not significantly change our conclusion. We believe that the stress fluctuation at low strain region was mainly a result of significant residual stresses at the low temperature due to the quench process of our metallic glasses. We have found that is possible to almost entirely remove these stress fluctuations by first applying small prestress to the material to presumably relax the residual stresses and we describe this pre-stress procedure in Supplementary Information (SI). Our observations strikingly resemble those of a crystal at low temperatures in that there is a fast, inertial motion in which the particles move ballistically, followed by caging after a timescale on the order of 0.1 ps in these materials [51]. At very long times, there is the hint that the particles are entering a regime in which they are no longer localized after the material has been sufficiently deformed, but this time regime is difficult to access by simulation at low strain values (We return to a consideration of this long-time regime below.). Importantly, $<r^2(t)>$ plateaus after the 'caging time' $t_{cage}$, a time that defines the Debye-Waller parameter, $<r^2((t_{cage})> \equiv <u^2>$. In the inset of Fig. 3(a), we see that this quantity varies linearly with $T$ in the glass state, as normally observed for $<u^2>$ in crystalline materials at $T$ much lower than the melting temperature [17]. The mobility of the particles in the metallic glass, defined by the mean amplitude of atomic motion, clearly depends on $T$ in a similar fashion in both crystalline and glass 'solids'. Corresponding to the plateau in $<r^2(t)>$, there is a plateau in the self-intermediate scattering function $F_s(q,t)$ having indefinite persistence in the time in the absence of deformation, signaling that the metallic glass is in a *non-ergodic state* (i.e., the density-density autocorrelation function does not decay to zero at long times). Appropriately, the magnitude of the plateau in $F_s(q,t)$ is often termed the 'non-ergodic parameter', which provides a quantitative measure of the 'degree of relaxation' that can occur in the glass state. As we shall discuss below, the α-relaxation time $\tau_\alpha$, which is characteristic relaxation time of cooled liquids can be taken to be *infinite.* Upon sufficient deformation, this relaxation process re-emerges so that $F_s(q,t)$ appears to decay as in a cooled liquid. We investigate this long time regime briefly below.

To quantify this transition in the dynamics, we examine how $<r^2(t)>$ is altered by a large deformation of the material. In Fig. 3(b), we apply a range of strains up to $\varepsilon = 8.9$ % for material with $h = 300$ Å at $T = 50$ K, with each curve calculated within $\varepsilon \approx 2$ % strain interval. The dynamics is unaffected in the short time regime, which is completely controlled by the particle kinetic energy, but at long times we observe a transition to delocalized behaviour at a progressively shorter time with increasing strain. The trend is again very much like what we would expect to see in crystalline materials at low *T* and cooled liquids upon raising *T*, although it is not clear that the dynamics is diffusive beyond the caged regime in the strained material. We briefly discuss this longer time dynamics below.

While the plateau value of $<r^2(t)>$ does not change greatly with strain in Fig. 3(b), a small variation in $<u^2>$ with $\varepsilon$ can be seen upon enlarging the scale in the inset of Fig. 3(b). The change of $<u^2>$ relative to its value at vanishing strain $<u^2(\varepsilon = 0)>$ and normalized by the square of the average particle distance, $\sigma$, $\delta<u^2> \equiv (<u^2> - <u^2(\varepsilon = 0)>) / \sigma^2$, increases nearly linearly with $\varepsilon$ over the large range of strain considered in our simulations. The magnitude of $<u^2>$ then depends on both a thermodynamic contribution $<u^2(\varepsilon = 0)>$ and a contribution arising from the applied strain, $\delta<u^2>$.

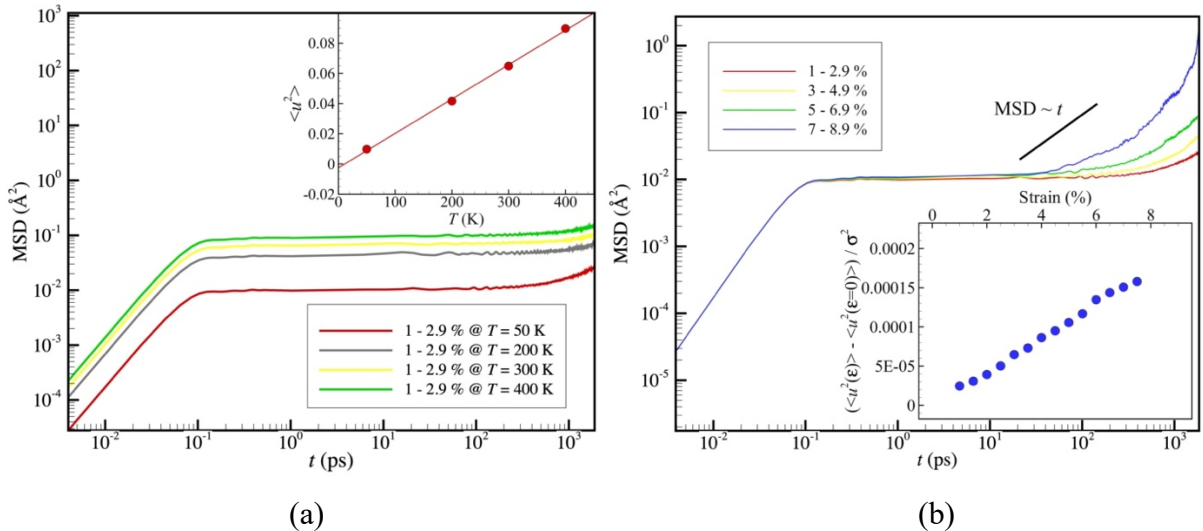

(a)                    (b)

**Figure 3.** Qunatification of atomic displacement in the glass state and particle caging. (a) Average mean square displacement $<r^2(t)>$ of all the atoms at different temperatures in the elastic regime. Inset shows that $<u^2>$ exhibits a linear relationship with *T*. (b) Average mean square displacement $<r^2(t)>$ of all the atoms at different strain levels. Inset shows the normalized $<u^2>$ as a function of applied strain.

In a previous study of shear in GF liquids [18], it was found that the structural relaxation time from the intermediate scattering function $F_s(q, t)$ approaches the caging time $t_{cage}$ at high rates of deformation and it is natural to consider whether a similar trend applies when the metallic glass is deformed. One difficulty in pursuing this question is that the definition of 'relaxation time' is not obvious in the metallic glass state. In experimental studies of metallic glasses, the Johari-Goldstein relaxation process is widely recognized as being the primary relaxation process in the glass state, but the direct investigation of this relaxation process involves the simulation or measurement investigation of mechanical relaxation processes at very low temperatures. In recent simulation studies of an Al-Sm metallic glass [21,22], however, we have shown that the Johari-Goldstein relaxation process essentially coincides with a specific measure of dynamic heterogeneity, the average lifetime $\tau_M$ of mobile particle clusters. This is a quantity more readily investigated by simulation and, conveniently, does not require the computation of mechanical properties of the material such as the stress autocorrelation function. We have previously investigated the distribution of the 'mobile particle clusters' in Zr-Cu metallic glass-forming liquids in the liquid regime, [27,52] and we next consider whether $\tau_M$ exists in the glass state and whether this quantity can serve as a relaxation time in the glass state, as in the Al-Sm system.

We should note here that the emergence of particle clusters having excessively high and low mobility in relation to expectations of Brownian motion is a general feature of GF liquids and this aspect of GF liquids, along with methodologies for identifying such particles, their distribution of size, average size, fractal geometry, lifetime, etc. have been discussed at length in previous publications on both metallic glass and polymeric GF liquids. [52,53] The reader is referred to this previous works devoted to this topic for the requisite technical discussion of these 'mobile' and 'immobile' particle clusters, which together define the dynamically heterogeneous nature of GF liquids. We find the same clusters arise in the glass state.

In Figure 4, we examine the average mass of the mobile particle clusters as a function of time $t$ using the same method of defining the mobile particles as described before [22] except that we consider a range of strain values between 3.0 % to 8.0 % in the glass state rather than a range of temperatures in liquid state. In the present analysis, we define 'mobile particles' as corresponding to the top 2.75 % the most mobile particles of the system at any point of time (see SI for the details on how we uniquely determine this cut-off value following the suggestion of Starr et al. [53]) As in our previous simulations in the 'liquid' regime above $T_g$ [52], the average mass of the mobile clusters generally exhibits a maximum at a characteristic time, $\tau_M$, which defines the 'mobile cluster lifetime'. We further see that increasing strain significantly decreases $\tau_M$, but the cluster size *increases*. In our previous study of $\tau_M$ in the liquid regime, we also found that $\tau_M$ was approximately equal the non-Gaussian parameter, a common measure of dynamic heterogeneity. The non-Gaussian parameter, $\alpha_2(t) = \frac{3<r^4(t)>}{5<r^2(t)>^2} - 1$ , measures the deviation of the displacement dynamics from a Gaussian, i.e., $\alpha_2(t)$ is defined to equals zero if the displacement distribution is Gaussian. The particle displacement distribution is Gaussian in fluids where the fluid dynamics is highly chaotic and the theory of Brownian motion applies, but this distribution also arises in crystalline materials at low temperatures where the particles are perfectly localized in their potential wells. The distribution function also becomes Gaussian in the short time inertial dynamics regime as a consequence of the Maxwell-Boltzmann distribution of the particle velocities for materials in equilibrium and the near ballistic nature of the particle motion in this short time regime. The magnitude of this distribution can thus provide information about whether the system is in at least local equilibrium or whether particle motion is localized, or a diffusion process reasonably modeled by Brownian motion. Correspondingly, we show the variation of $\alpha_2(t)$ as a function of time in Figure 5(a) for a range of strains, and in the inset of this figure we see that $\tau_M$ exhibits a nearly linear relationship with $t^*$ (the time at which the non-Gaussian parameter peaks) in the glass state. Recently, Puosi et al. also suggest a strong correlation between the peak time in the non-Gaussian parameter $t^*$ and JG β-relaxation in a model polymer glass-forming liquid using molecular dynamics simulation [54] so that this relation seems to have some generality, even if it is not universal. This suggests that this measure of 'dynamic heterogeneity' is common to both cooled liquids and the glass states and derives from the existence of mobile particle clusters. We note that mobile particle clusters having a lifetime $\tau_M$ also arise in heated crystalline materials [51] so this definition should apply just as well for crystalline materials both under quiescent or deformation conditions.

We have previously shown that the dynamics of the mobile particle clusters and the JG relaxation process in an Al-Sm metallic glass involves an intermittent particle 'jump' process. By 'intermittent', we mean that the jumping events can be described by a universal distribution with a power-law tail. Relaxation processes exhibiting this type of temporal heterogeneity exhibit non-exponential relaxation and the application of a renewal theory to relaxation in material systems indicates that relaxation in systems with this type of intermittency should generally be described by the Mittag-Leffler family of functions, which in the frequency domain is known as the Cole-Cole relaxation function [55,56]. This type of relaxation is commonly observed for the JG $\beta$-relaxation process in glasses [57,58], including metallic glasses when the alpha and beta relaxation processes are well-separated [59]. Mittag-Leffler relaxation leads to power-law stress relaxation at long times rather than the exponential or stretched exponential function (or the Havriliak-Negami function in the frequency domain [55,56]), characteristic of the alpha relaxation in the liquid regime or to power-law creep deformation under applied constant stress [60].

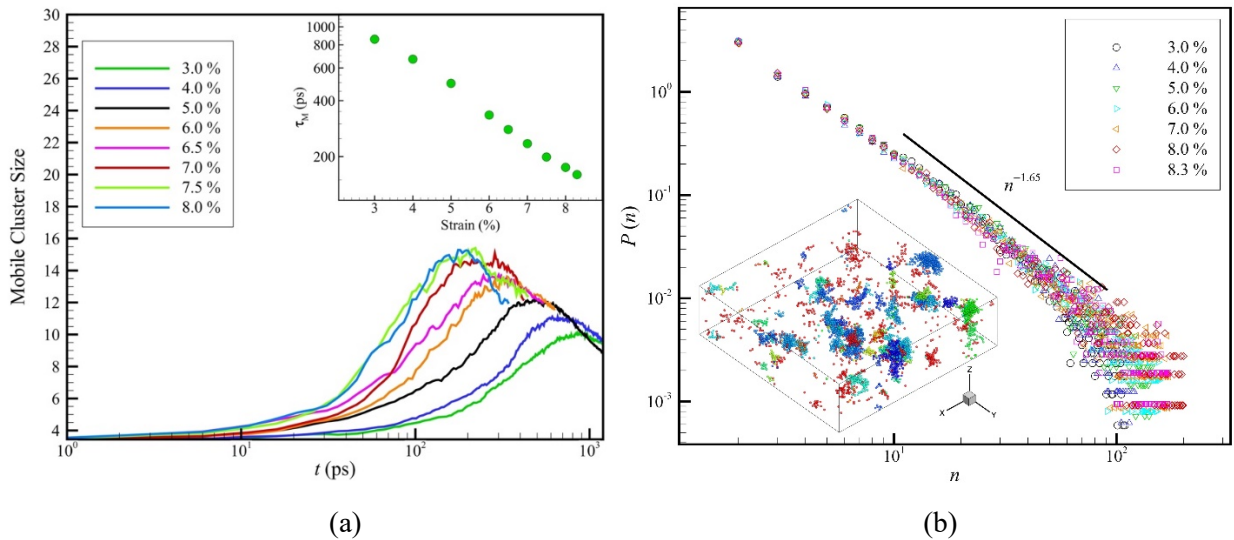

(a)                                                    (b)

**Figure 4.** Evolution of mass of mobile particle clusters in time, their lifetime with strain and their size distribution. (a) Mobile cluster size as a function of time at different strain levels for $T = 50$ K and $h = 300$ Å. (b) The mobile particle probability distribution function at different strains. The inset shows the mobile cluster configuration at 7.0 % with a size of $20 \times 20 \times 6$ nm. The evolution of the average size of the mobile particles and their size distribution is evidently similar to previous observations in the liquid regime, but strain clearly alters their size and average 'lifetime', determined by the time at which their average mass peaks. The size distribution of the mobile particles is nearly the same in the interior and interfacial regions, and the fractal geometry of the clusters in these two regions are almost identical, so there is apparently no essential difference in the geometry of the clusters in these regions. We have included a comparison of these distributions in Supplentatary Information, along with a quantification of the dynamic clusters, which are a good candidate for 'shear transformation zones'.

We suggest that the Cole-Cole relaxation function is *universal* for the Johari-Goldstein $\beta$-relaxation process observed in mechanical and dielectric measurements in glass materials. Power-law creep is a nearly universal property of diverse solid materials, encompassing glass, polycrystalline and crystalline materials having many different types of chemical composition (polymeric, metallic and ionic materials) [61–70]. The 'creep exponent' describing the progressive power-law change in the material dimensions under constant applied stress is often observed to be near 1/3 ('Andrade creep'), a value that Douglas and Hubbard have interpreted before as arising from polymeric chains of locally icosahedrally-packed particles in many glass-forming materials[55].

Although the α-relaxation process is inaccessible in simulations of both crystalline and glass materials at low temperatures because of its extremely long timescale or complete lack of existence, we may consider whether $\tau_M$ may serve as a 'relaxation time' appropriate for the glass state. We show in Figure 5(b) that the strain dependence of $\tau_M$ can be brought into an apparent universal reduced form in which $\tau_M$ approaches $t_{cage}$ corresponding to the limit of a nearly 'perfect fluid' as the strain is increased. In particular, $\tau_M$ is well-described by the scaling relation,

$$(\tau_M - t_{cage}) / t_{cage} \sim A \left[ (\varepsilon_c - \varepsilon) / \varepsilon_c \right]^{\delta} \quad , \quad A = 7.0 \quad , \delta = 1.37 \tag{2}$$

where the 'critical' strain, $\varepsilon_c = 8.3\%$ at which this limit is achieved corresponding to the point of shear band formation. This scaling mirrors the approach of $\tau_\alpha$ to $t_{cage}$ in the liquid regime for liquids subjected to increasing deformation [18].

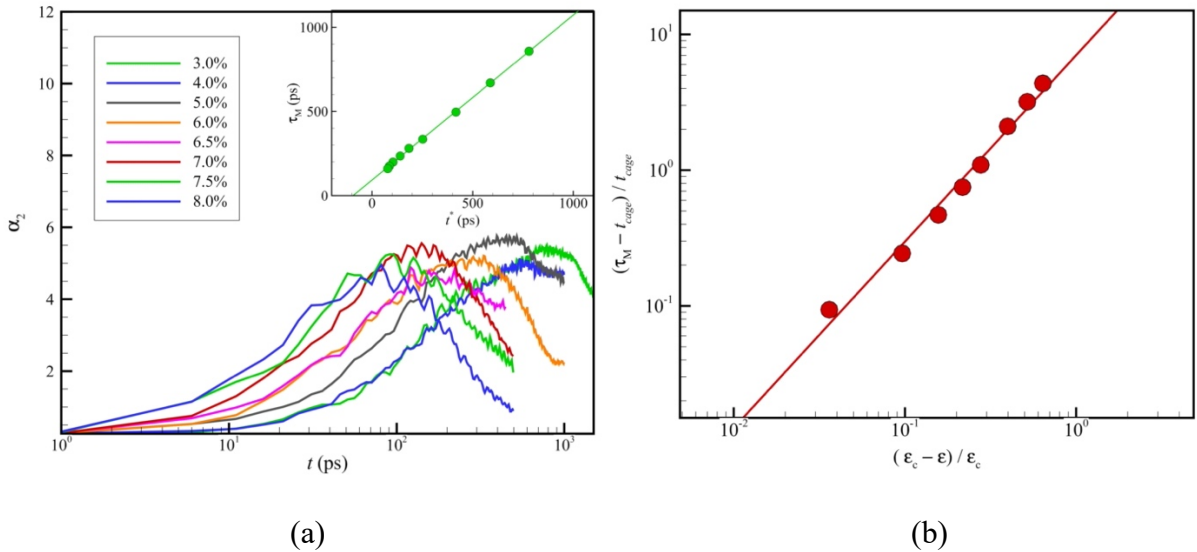

(a)                                                          (b)

**Figure 5.** Quantification of non-Gaussian parameter and its relation between its peak value and $\tau_M$. (a) Non-Gaussian parameter as a function of time at different strain levels. The inset shows the time $t^*$ at peak in non-Gaussian parameter exhibits a linear relationship with mobile particle lifetime $\tau_M$ at different strain levels. (b) Scaling relation between mobile particle lifetime $\tau_M$ and strain.

Recently, Giuntoli et al. [18] have emphasized that the approach of the relaxation time $\tau_\alpha$ of the fluid to $t_{cage}$ corresponds to the point at which the α-relaxation time merges with the fast $\beta$-relaxation time and that this condition corresponds to a flow instability condition because this condition implies that relaxation of the fluid by momentum diffusion (More precisely, the kinematic viscosity $\nu$, the ratio of the shear viscosity to the fluid density, is the momentum diffusion coefficient in hydrodynamic theory, but under the normal condensed fluid conditions in which the fluid density is nearly constant we may think of the shear viscosity as being proportional to the momentum diffusion coefficient.) becomes essentially suppressed at this point, i.e., the fluid can be viewed as roughly equivalent to an inviscid fluid from a modeling standpoint. 'Turbulent' or 'chaotic' flow is common observed in fluid flow having such exceptionally low viscosities [71], and, although there is no generally accepted theory of the transition to turbulent fluid flow, this tendency can be qualitatively understood from recent computational investigations of particle motion in idealized inviscid fluids, where it has been observed that the motion of a solid body through an incompressible, inviscid fluid, moving irrotationally, and otherwise at rest, is inherently chaotic [72]. It is notable that such turbulent motion has been observed in flowing polymeric liquids even under conditions in which the inertial energy of the fluid flow, as quantified by the Reynolds number, is small [73–75]. Simulations have recently shown that shear band formation after sufficiently large deformation of our metallic glass leads to the onset of hierarchical vortex formation, transient super diffusive motion, and other features common turbulent fluids (see discussion below.) This phenomenon in glass materials appears to be another variety of 'elastic turbulence'. Taylor [76] long ago stressed the analogy between turbulent flow in liquids and the heterogeneous flow expected to arise in association with the deformation of polycrystalline materials. Our results below support this intuition.

In order to quantify the emergent dynamic heterogeneity in our strained MG material, we next consider an apparently general relation between $<u^2>$ and the glassy shear modulus $G$ of the material obtained for the stress-strain relation in the limit of linear deformation. This relation will enable us to relate the shear modulus of the material over the entire accessible temperature range (200 K < $T$ < 1000 K) to $<u^2>$ of the material as a whole, and this also allows us to make a map of the local stiffness as a function of strain that will enable us to gain significant insight into the pre-transitional softening of the MG material leading up to the formation of SBs.

Our neglect of any consideration of α-relaxation in our simulations at $T$ well below the glass transition temperature, estimated to be $T_g$ = 817 K for the present metallic glass [27], deserves some explanation. In previous work on the same MG based on the same potential as in the present paper, we found that we could predict the α-relaxation time $\tau_\alpha$ without free parameters from simulation estimates of $<u^2>$ in the liquid state regime above $T_g$. In particular, the structural relaxation in the viscous liquid regime is predicted by the localization model to equal,

$$\tau_\alpha(T) = \tau_\alpha(T_A) \exp[(<u^2(T_A)> / <u^2(T)>)^{3/2} - 1] \tag{3}$$

where $T_A$ is an onset temperature for non-Arrhenius dynamics in the structural relaxation time and the rate of atomic diffusion. We also found in this former work that $<u^2>$ varied nearly linearly with $T$ at low temperatures below the crossover temperature $T_c$ = 1055 K, separating high and low temperature regimes of the glass-forming liquid [27,77].

The temperature variation of $<u^2>$ also serves to define two well-defined temperature regimes governing the glass state. In particular, $<u^2>$ estimates in the liquid regime formally extrapolate to 0 at a finite temperature $T_o$ = 711 K, the *same* temperature at which the structural relaxation time $\tau_\alpha$ diverges in corresponding fits of the Vogel-Fulcher-Tammann [78] estimates of $\tau_\alpha$ [27] (We emphasize that this extrapolation does not mean that the relaxation time actually diverges at $T_o$). The $T$ regime well below the glass transition, defines the 'low temperature glass state', the $T$ regime that is the focus of the present paper. Our estimates above of $<u^2>$ in the low $T$ glass state indicate that this quantity is positive, but $\tau_\alpha$ can nonetheless be reasonably taken as being 'infinite' for all practical purposes. Specifically, the localization model [27] for $\tau_\alpha$ that quantitatively describes in the $T$ range above $T_g$ indicates that the value $<u^2> = 0.0098$ Å$^2$ found at $T = 50$ K for our metallic glass implies a value of $\tau_\alpha \sim (10^{1153}$ years), which is quite large even in comparison with the estimated lifetime of the universe $\tau_{univ}$ indicated by the Lambda-CDM Concordance model [79], i.e., the time elapsed since the Big Bang, $\tau_{univ} \sim$ O($1.4 \times 10^9$ years). It seems safe to conclude that 'viscous relaxation' and $\tau_\alpha$ are not relevant to describing relaxation in low temperature glasses, and, by default, we define the relaxation time in terms of the Johari-Goldstein relaxation process, which we have found can be estimated from the lifetime of the mobile particles $\tau_M$ and $t^*$, as we discuss above.

In our previous paper, we found that diffusion still exists in the glass state, albeit its magnitude is relatively small, and that the diffusion coefficient $D$ scales as, $D / T \sim 1/ \tau_M$ [21,22] so that the rate of diffusion in the glass state is directly linked to $\tau_M$. Unfortunately, the determination of $D$ in the glass regime in the low regime that we study requires prohibitively long simulations. The rapid growth of $\tau_\alpha$ to astronomical time at low $T$ means that the plateau in Fig. 3 persists to essentially 'infinite' time in the low temperature glass state and there is evidence from both experiment [80] and simulation [81–83] that a true finite equilibrium (zero frequency) shear modulus can emerge in the glass state. Note that this analysis suggests that there is a separate high temperature glass regime for $T_o < T < T_g$ in which the Zr-Cu alloy exhibits a solid-like response over an appreciable timescale, but the material still exhibits a finite $\tau_\alpha$ so that the material exhibits viscoelastic response that is similar to a fluid over long timescales. This is the viscous 'plastic deformation' domain [84]. We have also argued that the relationship between the relaxation time $\tau_\alpha$ and $<u^2>$ derives from the weakly chaotic nature of GF and complex fluids in which this relation holds. Correspondingly, we may anticipate by this same reasoning that this relation to *breakdown completely* at low $T$ as the material undergoes an ergodic to non-ergodic transition upon cooling into the low temperature glass state. [85,86]

The decay of the intermediate scattering function to a plateau at long times as in a crystal at finite temperature, also implies that the material should exhibit a corresponding plateau in the shear stress relaxation function $G_p$ corresponding to a finite zero frequency shear modulus. The emergence of a finite 'equilibrium shear modulus' $G$ in the glass state has been discussed in numerous recent simulations, theoretical and experimental studies of glass-forming materials. [80–82,87–91] The localization of particles by surrounding particles is also signaled by a drop in the communal entropy of the fluid, a phenomenon observed in both materials that solidify into crystalline and non-crystalline solids. [92] This type of transition has been discussed as a characteristic feature of jamming transitions [93] and the same arguments seem to apply to the onset of the glass state. Apart from the breakdown of Eq. (3) relating $\tau_\alpha$ to $<u^2>$, and the lack of even the existence of a measurable $\tau_\alpha$, [94,95] this transition also implies the breakdown of ordinary thermodynamics. While the material still explores its phase space in a limited way in the 'glass' regime, we may expect the emergence of a generalized non-extensive thermodynamics under these circumstances, as illustrated by model computations dynamical systems at the edge of chaotic behaviour [96,97]. In particular, we may expect new limit theorems associated with non-extensive thermodynamics to emerge under these conditions [98], corresponding to a distribution function for particle displacements that greatly differs from the Gaussian functional form, which arises in the liquid regime along with a mean square displacement increasing linearly in time, as consequence of the limit theorem associated with the strongly chaotic nature of the fluid dynamics at elevated temperatures. We indeed see evidence of transient super-diffusion and a highly non-Gaussian distribution function for atomic displacement probability emerge that is consistent with Tsallis statistics [98], features that are observed in a wide variety of 'turbulent' systems. This type of highly non-Gaussian displacement distribution has been observed before phenomenologically in simulations of creep in a Zr-Cu metallic glass material in its glass state [99].

### C. $<u^2>$ as a Measure of Bulk and Local Material Stiffness

To address the question of local elasticity evolution with time, position and strain, an appropriate and computationally feasible measure of local material 'stiffness' is required. In addition to being a measure of 'mobility', we argue below that $<u^2>$ can also be interpreted physically as a measure of material stiffness. The Debye-Waller parameter $<u^2>$ has the dual advantage of being defined at an atomic scale and of being readily measurable by X-ray, inelastic neutron scattering, and other scattering techniques [100–102], although spatially resolved measurements at a nanoscale are not currently possible. Simulation studies on both polymeric GF liquids [26] and metallic GF liquids [27], have established a linear scaling relationship between shear modulus $G$ and $k_B T / <u^2>$, where $k_B T$ is the thermal energy and we consider the validity of this relation in the present material, first in the bulk material and then as providing a measure of local material stiffness.

To test this scaling relation linking $G$ and $<u^2>$ in the glass state in which our shear banding simulations are performed, we first determined $G$ by applying constant shear rate along one direction on $Cu_{64}Zr_{36}$ alloy bulk samples at different $T$ ranging from 200 K to 1000 K, and calculated stress-strain curves at different $T$ in Fig. 6 (a) to obtain the estimates of $G$ as a function of $T$ shown in Fig. 6 (b). The inset to Fig. 6 (b) shows the linear correlation between $G$ and $k_B T / <u^2>$ in $Cu_{64}Zr_{36}$ previously observed by Douglas et al. [27] at room temperature. This type of relation, although the specific functional form is somewhat different, has been discussed for many systems previously by Leporini and coworkers [31,32]. The relationship motivates defining $k_B T / <u^2>$ as a measure of *local* 'stiffness' that we will then consider in connection with the onset of shear band formation.

Before making a local 'map' of material stiffness based on $<u^2>$, we discuss the motivation for applying this scaling relation between $G$ and $<u^2>$ locally within the material and the limitations of this 'correspondence'. First, we note that the shear modulus $G$ is defined in the thermodynamic limit and that there is no obvious unique local counterpart of this material property. This situation provides us with some latitude with defining a local measure of material 'stiffness' that is consistent with the definition of $G$ in the thermodynamic limit. Under the circumstances, we think that it is best not to identify this local measure of 'stiffness' as being *exactly* the local shear modulus, however. Accordingly, we define the local stiffness $S$ by the relation,

$$S \approx S_o + S_l \, (k_B T) / \langle u^2 \rangle \, l \tag{4}$$

where $S_o$ and $S_l$ are fitted constants defined by the *macroscopic* scaling relation, and $S$ exactly reduces to the glassy modulus $G$. Leporini and coworkers [31,32] have discussed this phenomenological equation at length and have tried to interpret $G_0$, and the mysterious length scale $l$ arises in Eq (4) by dimensional consistency. A full understanding of $S_o$ and $S_l$ and $l$ remain an outstanding question that we do not concern ourselves here where we are interested only in obtaining a qualitative stiffness measure.

Notably, this same relation between 'stiffness' and $k_B T / <u^2>$ has often been reported to apply at an *atomic scale* in the interpretation of neutron and x-ray scattering measurements [103–111].

The motivation for taking $<u^2>$ as a measure of local molecular rigidity derives from the harmonic oscillator model, a molecular model of 'elasticity' in which the harmonic oscillator force constant $k_h$ equals $k_B T / <u^2>$ (see Appendix B of Ref. [112]). At an intermediate scale between atoms and bulk materials, this type of scaling relationship has also been tested affirmatively in molecular dynamics simulations of the shear rigidity of semi-flexible polymers where the effective polymer rigidity inferred from the persistence length, along with well-known estimates of the dependence of the persistence length of worm-like chains as a function of shear rigidity, again conforming with Eq. (4) (See Appendix B of [112]). We collectively infer from these numerous observations that we might quite reasonably interpret $k_B T / <u^2>$ as a local measure of material 'stiffness', although the literal identification of this quantity with a local shear modulus is questionable. To further establish the physical basis of this local 'stiffness metric', we consider an alternative measure of local stiffness and compare to the $<u^2>$ derived measure.

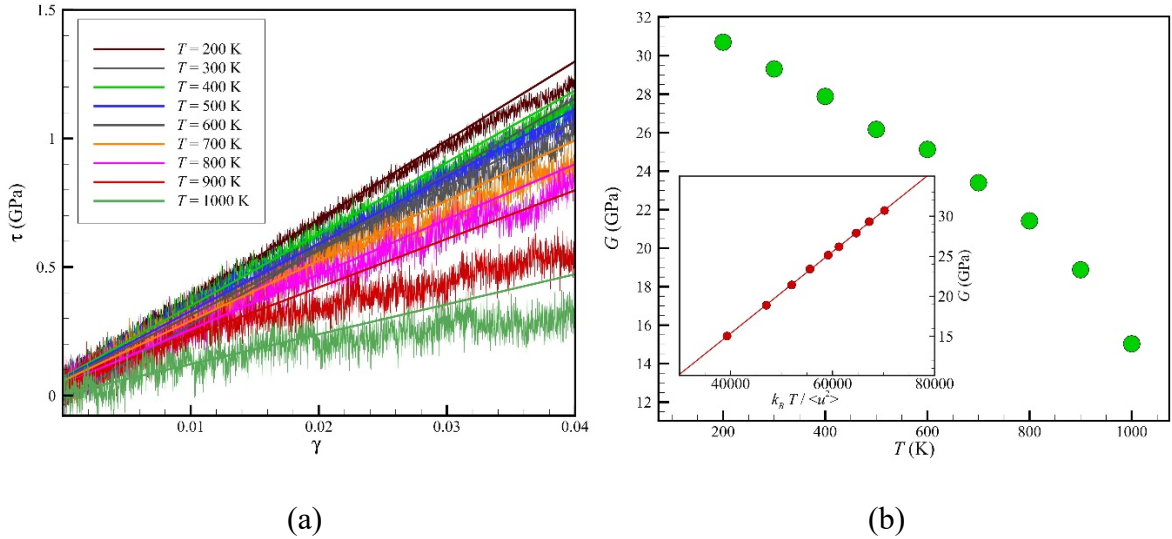

(a)                                                            (b)

**Figure 6.** Estimation of shear modulus as a function of temperature from deformation in the near elastic regime. (a) Stress-strain curve for $Cu_{64}Zr_{36}$ over a range of $T$ under shear deformation. (b) Shear modulus $G$ versus $T$ for the bulk material. Inset shows a linear relationship between $G$ and $k_B T / <u^2>$ for the bulk $Cu_{64}Zr_{36}$ metallic glass material.

As an alternative to taking $k_B T / <u^2>$ as a measure of local stiffness, we may independently estimate local material stiffness based on a consideration of *local* stresses and strains in the material at the atomic scale. In particular, we may define the 'local elastic constant' by a formal extension of its macroscopic definition as, $C_{11,i} = \Delta\sigma_i / \varepsilon_i$, where $\Delta\sigma_i$ has been taken as the von Mises stress on each particle $i$, and $\varepsilon_i$ is the von Mises strain of each particle [111,113]. Figure 7 shows the contour maps of the local elastic constant and local stiffness measure $k_B T / <u^2>$ for a $\varepsilon = 2.0$ % strain. In general, the correspondence between the colormaps describing the magnitudes of $C_{11}$ and $k_B T / <u^2>$ is reasonable, but not perfectly isomorphic. The average scale and variance of the fluctuations in these stiffness measures seem to hold very well, however. We take this as further evidence that it is reasonable to consider both $k_B T / <u^2>$ and $C_{11}$ as local stiffness measures, although we again emphasize that we must refrain from exactly identifying either $C_{11}$ or $k_B T / <u^2>$ with a local shear modulus.

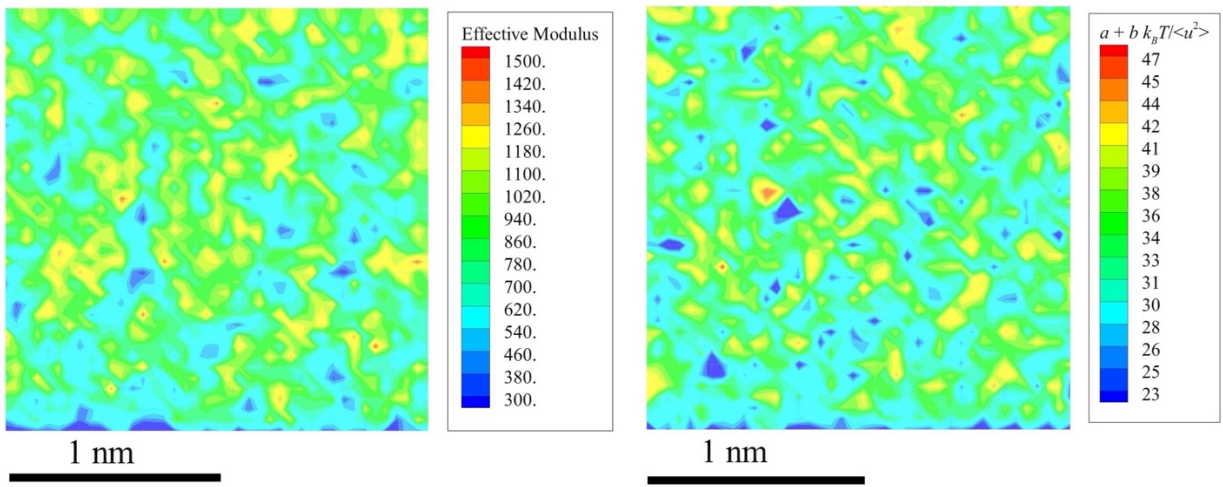

**Figure 7.** Contour maps of the local elastic constants ($C_{11}$) and local shear modulus ($k_B T / <u^2>$) at $\varepsilon = 2.0$ % of a 200 Å x 200 Å x 5 Å slab.

## D. *Local 'Softness' Evolution During Shear Band Formation*

As emphasized in the Introduction, It has long been suggested that the initiation of shear band formation should initiate in soft spots in the material in which large non-affine deformations are concentrated [114–117] and it is for this evident reason that we have defined a precise local measure of material stiffness to test this hypothesis.

If this interpretation of the mechanism underlying SB formation holds, then we may expect that the stiffness of the region forming the SB to be rather distinct from regions that remain in an ordinary glass state that is more similar to the undeformed bulk material than the SB region. The material also has an interfacial region in which the dynamics of the SB and bulk regions of the MG material are distinct from the material interior (See Fig. 8a) [28]. The SB interfacial regions are defined by the atoms located in the upper and lower surfaces area with a thickness of 15 Å and these regions are marked in Figure 8(a). The following analysis is based on the sample with $h = 300$ Å.

We then analyze how the stiffness $\mathcal{S}$ in the defined regions of Fig. 8a evolves as the material is deformed. In particular, we define $\zeta$ as the ratio of the number of atoms having a value of $k_B T / <u^2>$ less than the critical value to the total number of atoms in the designated region. In our analysis, the critical value of $k_B T / <u^2>$ is taken to be specifically 33 KJ/m$^2$, which corresponds to this stiffness measure at the crossover temperature $T_c$ at which previous molecular dynamics studies have established the onset of highly anharmonic 'liquid-like' motions in the interfacial regions of both crystalline and metallic glass materials [28]. The crossover temperature $T_c$ separates the high and low $T$ regimes of the glass-formation [27].

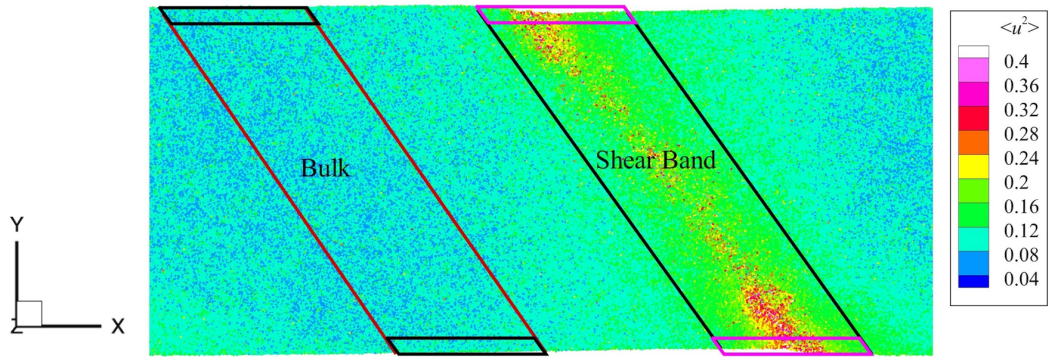

(a)

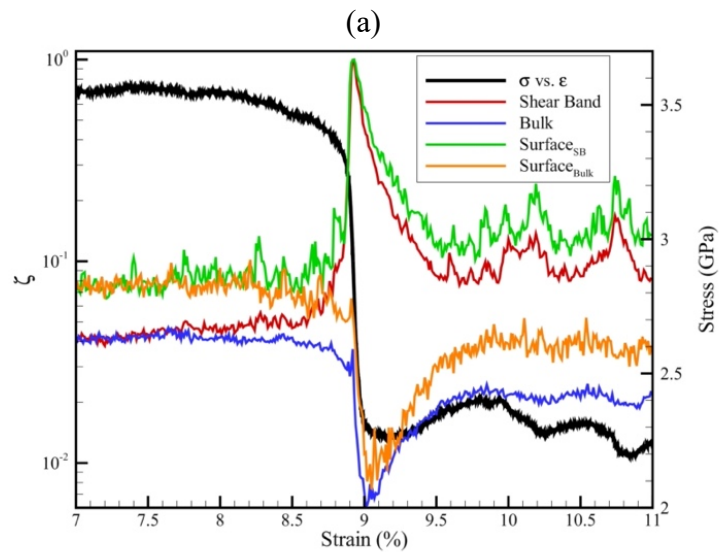

(b)

**Figure 8**. Shear band and bulk regions are defined in terms of local 'stiffness' and the evolution of stiffness under strain. (a) Distribution of $<u^2>$ in the sample, showing the shear band and bulk regions, as defined in the text. Interfacial SB and bulk regions having a thickness of approximately of 15 Å are indicated for both regions [118]. (b) The evolution of the fraction of softness in the strain range of 7.0 % to 11 %. Four areas were tested, SB region, bulk region, surfaces of SB region, and surfaces of the bulk region. The corresponding stress-strain curves are presented as the black line.

The black line in Figure 8(b) corresponds to the stress-strain curves for $\varepsilon$ between 7 % to 11 % and the sharp change in $\zeta$ indicates there is a critical strain region at which shear banding initiates. Apparently, the fractions of atoms in SB and bulk regions are almost identically near, but below the stress maximum defining the 'yield' of the material (see Fig. 1), i.e., a strain of 7.6 % at this $T$. Beyond this point, the soft regions in the SB apparently mediate the plastic deformation localized to the SB. A similar behaviour, but greatly amplified, is observed in the interfacial regions where the fluctuations in $\zeta$ are especially large. We next consider how the distribution of local stiffness, defined in terms of $<u^2>$, varies with deformation both deep within the MG material and in the interfacial region, where we find that the onset of SB formation corresponds to a 'critical' condition in which the stiffness within the material becomes equal to its value in the interfacial region in the absence of deformation. This instability condition was completely unanticipated.

### E. *Shear Band Formation as an Emerging Interface within the Metallic Glass Material*

The full development of a SB has one of its implications as emergence of an interface within the material, and we may therefore expect that the dynamics within the incipient SB region to progressively become similar to the dynamics of interfacial region of the material. We have recently quantified the interfacial dynamics of both metallic glass and crystalline materials based on modelling that emphasizes the gradient in $<u^2>$ in the interfacial region of these materials [28] so that the current analysis is a natural extension of this previous work. We first show that this approach of the elastic fluctuations in the interior of the incipient SB indeed approach that of the interfacial region and we follow this analysis by introducing a measure of the degree to which the atoms in incipient SB have been converted into a dynamical state consistent with the interfacial dynamics of the material.

Figure 9(a) shows the local stiffness $S$ distribution of the entire sample ($h = 300$ Å) at different strains before the maximum stress, with distribution on surface and interior as reference. The overall probability distribution of $S$ for the interior of the material evidently progressively approaches the distribution of $S$ of the interfacial region in the absence of strain as $\varepsilon$ increases from zero to the critical strain for SB formation. Similar behaviour has been observed in systems with $h$ = 90 Å, 150 Å.

We note that the average magnitude of $<u^2>$ within the region that ultimately forms a shear band changes from $<u^2>$ = 0.0098 Å² to a value $<u^2>$ = 0.0109 Å² at the onset of shear band formation. This change in $<u^2>$ allows us to estimate the change in the relative stiffness of SB region compared to the undeformed material, $\delta G \equiv [G(\text{SB}) - G(\varepsilon = 0)] / G(\varepsilon = 0)$. For the $h$ = 300 Å film at $T$ = 50 K, we find $\delta G$ = –0.12. The shear band regions are inherently softer than the undeformed material. A softening in the SB region has been reported experimentally in indentation studies of metallic glass materials [119].

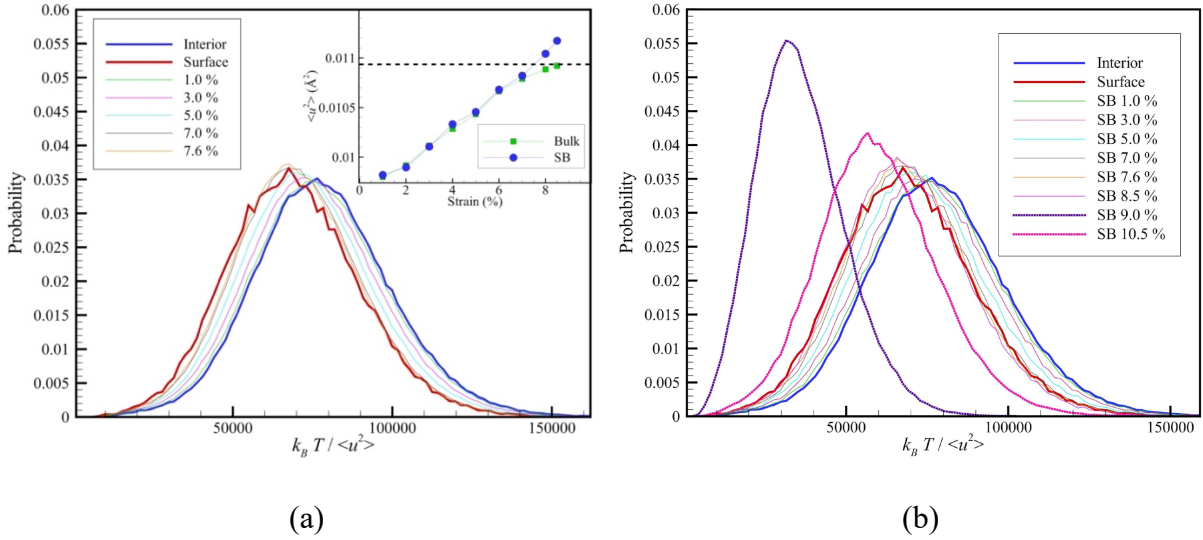

(a)                                      (b)

**Figure 9.** Distribution of the effective stiffness distribution in the interfacial regime and with the metallic glass material in the deformed and undeformed states for $h$ = 300 Å film at $T$ = 50 K. (a) Comparison of shear modulus distribution of entire sample at different strain rates before reaching maximum stress. Blue and red lines are the reference distribution at center and surface at zero strain. The dashed line in the inset is the $<u^2>$ of the interfacial region at the undeformed state. Apparently, the $<u^2>$ of the bulk and shear band approaches the value of $<u^2>$ at the interfacial in the absence of deformation, this convergence effect is illustrated in the inset figure. (b) Comparison of the estimated stiffness distribution within shear band at different degrees of strain with the stiffness distribution in the interfacial region ('surface') and the interior of the MG material ('interior') in the absence of strain.

We next examine the distribution of $S$ in the SB region in Fig. 9. As the strain approaches the 'yield' value of $\varepsilon$ = 7.6 %, at which a maximum in stress or 'yield' (see Fig. 1) occurs, the SB starts to form and the local stiffness ($S$) distribution essentially *coincides* with the interfacial $S$ distribution near the point of yield. The significance of the relative value of $<u^2>$ in the interior of the material to that on the boundary under undeformed conditions is even more apparent in the inset of Fig. 9a, where the magnitude of $<u^2>$ in the SB and the rest of the interior of MG material ('bulk') are nearly equal all the way up to the point of SB formation and yield. There is a clear suggestion from these observations that SB formation in our material reflects an instability in which the interior of the material has softened under deformation to such a degree that the stiffness is similar to the stiffness of the material prior to deformation. It apparently becomes energetically favorable for SBs to form when it becomes energetically favorable for the formation of interfaces within the material through local deformations arising within the material. As a practical implication, we note that the interfacial properties of metallic glass and other glass materials can be expected to greatly depend on how the material was fabricated so that the actual value of the critical deformation required for SB formation can then be expected to depend on exactly how the material was fabricated, especially with regard to fabrication processes what influence the interfacial properties of the material. This sensitivity of SB formation and yield to the surface 'states' has been demonstrated in recent studies of metallic glasses [120–122], consistent with our general finding that $<u^2>$ in the boundary of the MG material plays a crucial role in determining the onset condition for SB formation. Ivancic and Riggleman [123] have recently observed in a strong sensitivity of shear banding to defects in the interfacial region in simulations of a coarse-grained polymer material in conjunction with machine learning methods to analyze their data.

We point out that recent measurements on glass-forming liquids have indicated the dynamics of the interfacial region is largely 'decoupled' from the interior of the material [124] so that we expect the interfacial value of $<u^2>$ defining the onset of SB instability to be rather insensitive to film thickness. This expectation remains to be checked by simulation and measurement.

This instability condition reminds us of another interfacial instability condition seen in connection with the interfacial dynamics and the dynamics within both crystalline and glass-forming materials. Even in the absence of material deformation, the average value of $<u^2>$ in the interfacial region and interior of both crystalline and glass-forming materials extrapolate to a common value, defining the 'Tammann temperature' at which the interfacial region starts to acquire a greatly enhanced mobility, and interestingly, this temperature appears to correlate strongly with the glass transition of the material [29]. Notably, this convergence of $<u^2>$ values within the material occurs near $T_g$ and $T_m$ in glass-forming materials (See Fig. 2b of Mahmud et al. [28]) and the relevance of this phenomenon to the low temperature non-ergodic glass state is an open question.

Since we think that this condition is a related phenomenon to the SB instability that we observe, we found $<u^2>$ in the interfacial region and the interior of the material are seen to converge around 24 K. This is similar to our previous observations in the same metallic glass-forming liquids above $T_g$, where $<u^2>$ in the interfacial region and the interior of the material converge at a temperature close to glass transition temperature [28]. This 'premelting' or 'softening' would appear to be another interfacial instability associated with the lowered energetic cost of creating interfaces within the material and this instability appears to be of a very general nature.

The occurrence of the Tammann temperature in both crystalline and glass materials also makes us wonder whether a corresponding interfacial instability might arise in connection with the plastic deformation of crystalline materials. This is not the topic of the present paper, but we briefly note that both shear banding [125] and yield [126] are likewise observed in crystalline materials and the importance of interfacial mobility in the plastic deformation of small scale crystalline materials has also been emphasized [127–131]. In particular, the Zhu et al. [132] have noticed empirically that the barrier height for dislocation nucleation vanishes at a 'surface disordering temperature' (This temperature is typically about (1/2 to 2/3) $T_g$ and thus quite distinct from $T_g$.), and characteristic temperature is consistent with our definition of the Tammann temperature. It has also been noted [133] that the brittle-ductile transition for dislocation free crystals follows the empirical rule (2/3) $T_m$., which further supports the potential relevance of the Tammann temperature for the plasticity of crystalline materials.

The obvious implication of these observations is that it should be possible to engineer the surfaces of both crystalline and glass materials to impact their plastic deformation. This possibility was suggested for crystalline materials by Zhu et al. [132] and recently realized in practice by Shin et al.[134] by modifying the surfaces with coating the material using atomic layer deposition. Recently there has been great interest in the strengthening of glass materials by modifying their interfaces, so-called "chemical strengthening" [135] and there would appear to be great scope for interfacial engineering the properties of materials based on the material processing changes that impact interfacial mobility.

One of the interesting implications of the occurrence of a Tammann temperature in crystalline and amorphous solid materials is its effect on material yield (see Fig. 1). As $T$ approaches this characteristic temperature at which $<u^2>$ in the interfacial region approaches $<u^2>$ deep within the material interior we may expect the material to become inherently unstable to plastic deformation ('flow' in the colloquial sense) without any appearance of 'yield'. Consistent with this argument, many experimental studies on both metallic and polymeric glass-forming liquids have indicated a general tendency for the yield stress to approach 0 as the $T$ approaches the glass transition temperature of the material, a temperature that coincides closely with the Tammann temperature [84,136–138]. The same trend is observed in semi-crystalline polymers where the onset temperature is well below the melting temperature at which significant mobility emerges in these complex polycrystalline materials [84]. We remind the reader that this reasoning does not necessarily extend to the phenomenon of shear banding region since $\alpha$-relaxation remains prevalent for an appreciable temperature range below $T_g$. It is not clear that shear banding accompanies yield in this *high temperature glass regime*, i.e., $T_o \leq T \leq T_g$ where we interpret $T_o$ to be an ideal glass transition temperature at which the loss of ergodicity occurs. We may estimate $T_o$ as the temperature by extrapolating $<u^2>$ of the material to 0 at low temperatures and in the present MG material this characteristic temperature was estimated[27] to equal, $T_o = 711$ K.

The formation of SB is not the end of the story. Upon increasing the strain further there is another distinct region beyond material 'yield' in which the SB regions undergo various transformations and in which the dynamics within the SB region changes greatly from the undeformed material. Viewed from a stiffness map perspective, regions rich in 'soft' particles at $\varepsilon = 9.0$ % to regions rich in 'stiff' particles for $\varepsilon = 10.5$ % so that a kind of strain hardening apparently occurs in the SB region under post-yield conditions. The probability distribution of $\mathcal{S}$ also appears to move towards larger $\mathcal{S}$ values when $\varepsilon$ is progressively increased from $\varepsilon = 9.0$ % to $\varepsilon = 10.5$ %. After shear band formation initiates near $\varepsilon = 7.6$ %, there are non-trivial 'jerky' fluctuations in the stress-strain curve that arise from stress-induced particle movements that mediate large scale plastic deformation of the material. The local stiffness in the SB fluctuates greatly until the SB spreads throughout the sample. Below, we analyze this regime, where we find that the dynamics exhibits features similar to fluids undergoing a transition to turbulent flow, evidently corresponding to a kind of 'elastic turbulence'. The dynamics of this regime becomes much richer in this regime than before SB formation. Before briefly discussing this regime, however, we provide some further quantification of SB formation from a local stiffness perspective.

### F. Brief Discussion of the Dynamics in the Shear Band Region Beyond the Yield Condition

As might be expected from the discussion above, the dynamics of the MG material becomes much more complicated beyond the point of yield and associated SB formation. An examination of the particle motions in this regime indicates the occurrence of complex vortex motions that are reminiscent of turbulent fluids [139–142]. A contour map of local shear modulus $k_B T / <u^2>$ near shear band and displacement vector in the corresponding region exhibiting 'turbulent' were shown in SI. This vortex state is a highly complex phenomenon with important ramifications for the dynamics of strongly deformed amorphous solid materials and we briefly address some of the conspicuous features of this dynamical state based again on a dynamic heterogeneity perspective.

One interesting phenomenon occurring in turbulent media, such as the earth's atmosphere, is that the average particle displacement from an initial position has often been observed to occur in a super-ballistic fashion over appreciable space and time scales [143–145]. In particular, the mean square displacement has often been observed to increase as $t^3$ compared to $t^2$ of ballistic motion and $t$ for Brownian motion [145]. While there is still no rigorous hydrodynamic theory explaining this phenomenon, it is generally appreciated that this super-ballistic transport derives from highly correlated velocity fluctuations in the atmosphere that drive the particle motion. Above we noted the existence of vortex-like fluctuations in the shear band regions that are highly developed in the SB region beyond yield, and we correspondingly examined the $t$ dependence of the mean square atomic displacement $<r^2>$ for atoms in the 'turbulent' SB region. We find that $<r^2>$ increases very sharply with $t$ in this unstable regime with an apparent power that is very large, as illustrated in Figure 10. It is probably best to think of this stress-driven transport as reflecting the macroscopic deformation within the SB associated with incipient 'fracture' (See Fig. 1 b of Shrivastav et al. [34] where it is shown that there is a crossover back to diffusion at still longer times in their simulations of SB formation.) This type of crossover also arises at long times in passively driven particle displacement in turbulent fluids [145]. These brief observations offer only a small hint of the complexity of the dynamics within fully developed SBs.

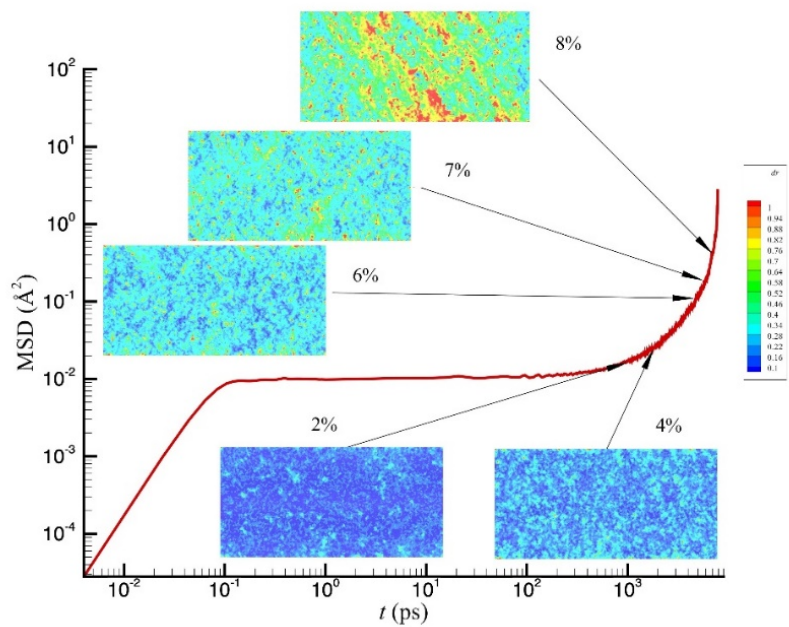

**Figure 10.** Atomic displacement and the evolution of stiffness heterogeneity. Mean squared displacement (compared with the initial state at $\varepsilon$ = 1%) in the sample with $h$ = 300 Å at $T$ = 50 K. Displacement distributions at $\varepsilon$ = (2, 4, 6, 7 and 8) % are shown in the graph. At 2 %, the sample shows non-localized displacement, but beyond about $\varepsilon$ = 6 %, areas with higher displacement marked with red color start to appear and areas apparently trigger relaxation in the surrounding environment.

We note that superficially similar vortex excitations, 'turbulence' and transient super-diffusion 'driven' by the vortices occur in a wide range of systems beyond ordinary turbulence and the elastic turbulence of flowing polymer melts and deformed metallic glass materials. This type of phenomenon is observed in flowing granular media [146–150], dusty plasmas [151,152], the movement of cells in living tissues [153–155] and 'active matter' in the form of living bacterial and alga suspensions [156–159]. Recently, there has been intense interest in understanding this generalized turbulence phenomenon [160]. Oyama and coworkers [149] and others [161,162] have shown in the context of flowing granular fluids that the vortices are comprised of clusters of particles moving in a string-like fashion and Liu et al. [118] have also shown this type of relationship between string-like motion and the vortices formed in deformed metallic glasses. We have confirmed that this hierarchical structure of the large-scale vortex patterns is composed of a finer structure involving cooperative atom exchange events and we will report our findings on this complex phenomenon in a separate paper devoted to this topic. This type of hierarchical structure has also been observed in GF liquids [38].

It should be apparent by now that glasses, GF liquids, and highly deformed materials generally, exhibit a rich dynamic that reflects the dynamic heterogeneity intrinsic to these materials, even in highly deformed crystalline materials [125,126,163], which embody arguably most forms of condensed matter. The dynamic formation and disintegration of structures in these materials (intermittency) imply that the material properties correspondingly exhibit large fluctuations, offering valuable information about the scale and geometry of the self-assembly processes underlying this spontaneous clustering arising from many-body dynamics. In practical terms, this means that basic properties such as potential energy become highly 'noisy' [163–165], and exhibit long range correlations in the form of colored noise and stress fluctuations and particle displacements exhibiting quake-like events having an exponential distribution in intensities and power-law distribution of occurrence are naturally observed in this type of system, even under equilibrium conditions. We have discussed this phenomenon at length in connection to the dynamics of the mobile particle clusters and Johari-Goldstein relaxation process in our previous work focussing on a Al-Sm metallic glass [21] and we have investigated this same rather generic phenomena in a variety of other materials such as interfacial dynamics of Ni nanoparticles [13], the interfacial dynamics of bulk crystalline Ni and ice [166–168] and internal dynamics of proteins [103]. Many others have studied this type of 'quake' phenomenon in deformed glass materials, [169,170] where clusters with geometric properties consistent with our mobile particle clusters were inferred to be the origin of this phenomenon. We also leave this interesting aspect of the highly deformed glass state for a future investigation. Here we point out that studies of noise offer much valuable information about the dynamic heterogeneity in this class of materials and this noise itself is of intrinsic interest in biology and various fields of material science where it is functional in relation to biological sensing and the catastrophic failure of materials [171–176].

Many studies have previously shown that the large fluctuations in the various types of systems exhibiting turbulent dynamics lead to transient anomalous diffusion in which the fluctuations accelerate the movement of particles in the system. We next show that the van Hove function describing atomic displacement in a highly deformed metallic glass exhibits rather typical behavior for this type of elastically turbulent system in the regime in which shear bands have formed. Given the potential practical importance of this phenomenon and the lack of previous quantification of this phenomenon in deformed metallic glass materials, we consider the particle has displaced a distance $r$ from its initial position (taken at the origin) at time $t$, i.e., the van correlation function $G_s(r,t)$ in liquid state dynamics jargon in Fig. 11 at the caging time at which $<u^2>$ is defined and at the time $\tau_M$ at which the mobile particles are defined, a time that also essentially coincides in our material with the peak time of the non-Gaussian parameter. The displacement at the caging time is nearly exactly described by a Gaussian function (see inset to Fig. 11) where $<u^2>$ defines the atomic displacement on a ps timescale defining the caging time, but $G_s(r,t)$ at $\tau_M$ is clearly highly non-Gaussian. The 'fat tail' describing the particle displacement on intermediate is a rather universal characteristic of systems exhibiting the highly collective dynamics, as observed in the present material, and 'turbulent' systems broadly [177]. The long tail in this type of distribution has often been described by stretched exponential distributions [177,178], as in Richardson's original model of passive transport of particles in a turbulent medium [142,143], and sometimes this distribution is approximated as being exponential for simplicity. [157,179,180]

As noted before, we may expect the deformed metallic glass to be in a non-equilibrium state in which ordinary equilibrium thermodynamics does not necessarily apply. As in the case of ordinary turbulence, we follow precedent in this type of situation and fit $G_s(r,t)$ to a form expected from non-extensive thermodynamics, the thermodynamics appropriate in which the system is not quite ergodic, but not really periodic or regular in its dynamics either [96,97,148]. The solid line in Fig. 12 shows a fit to the general Tsallis distribution, as prescribed by Eq. (15) of Beck [98]. We see that this distribution function, which has been shown to fit probability distribution functions for turbulent fluids and a variety of non-equilibrium systems, is rather well. The Tsallis theory [98,181] does not provide a specific prediction of the characteristic exponents describing the non-Gaussian distribution from first principles, but we may conclude again that the basic phenomenology that we observe for SB region is typical of 'turbulent' materials.

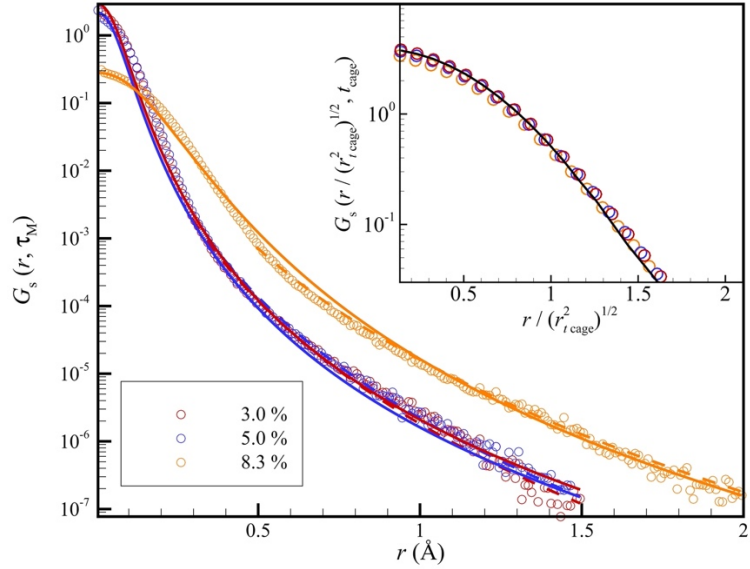

**Figure 11.** Van Hove correlation function with $\tau_M$ at $\varepsilon = 3.0$ %, 5.0%, and 8.3 %. The tail of the van Hove function is fit to the generalized Tsallis distribution [98,178] (solid lines), $G_s(r,t) \sim [1 - \beta(1-q)\,r^\delta]^{1/(1-q)}$ where $q = 1.39$ and $\delta = 2.31$ for $\varepsilon = 3.0$ % and 5.0%, and $q = 1.32$ and $\delta = 2.2$ for $\varepsilon = 8.3$ %, respectively. The dashed curves indicate the apparent 'stretched exponential' tails of $G_s(r,t)$ where the stretching exponent $\beta$ values are noted in the text. The inset shows the van Hove function at the same strains at the caging time, 1 ps where we have introduced a reduced displacement distance, $r / <u^2>^{1/2}$, and where the solid line is a Gaussian function. The particle displacements at this short time are remarkably similar to those found in perfect crystals at low temperatures and are typical generally for materials in equilibrium.

We note that the tail of $G_s(r,t)$ can also be reasonably well-described phenomenologically by a stretched exponential of index $\beta$, which jumps in its value from about $\beta = 0.2$ below the regime of SB formation to a value $\beta = 0.25$ near the point of SB formation, a phenomenon that reflects the sharp transition to a turbulent state in which the generation of a hierarchy of vortices that greatly influences the transport of particles on intermediate time and space scales. At very long timescales, we expect $G_s(r,t)$ for particle displacement to return to a Gaussian form, as seen also in ordinary turbulent, granular fluids, and nearly jammed fluids [145,146,182,183], and also observed in the context of simulations of deformed metallic glasses by Shrivastav et al. [34]

**Conclusions**

In the current study, we investigated the deformation mode in $Cu_{64}Zr_{36}$ alloys with different thicknesses and under different temperatures. A different deformation mode has been observed in these systems, which leads to a question: How to characterize the localized and non-localized deformation? Von Mises strains were implemented to monitor the formation of shear band. We found that the shear band gradually builds up inside the system and undergoes soften and harden process as we increase the strain. Eventually, one dominant shear band develops when the localized deformation mode occurs. Since the temperature when the deformation happens is relatively low, atomic diffusion is extremely slow and other processes should be expected to happen to influence the structural relaxation.

1.   We first considered the average mean square displacement of our metallic glass deep in the glass state and found that dynamics in this regime seemed to greatly resemble that of crystalline materials at a temperature well below their meting point. In particular, the mean square particle displacement $<u^2>$ after a well-defined caging time, the 'Debye-Waller factor', is relatively small and increases nearly linearly with temperature. No detectable evolution of $<u^2>$ was observed in our metallic glass material, but we cannot exclude such a possibility on timescales longer than we could simulate.

2.   The intermediate scattering function does not decay from the plateau determined by $<u^2>$ so no finite $\alpha$-relaxation time appears to exist in the glass state. Experimental studies have established that the Johari-Goldstein relaxation time ($\tau_{JG}$) exists in the glass state, however, and adopt criteria developed in previous papers investigating the Johari-Goldstein relaxation process in which the Johari-Goldstein relaxation time is identified with the mobile particle lifetime, $\tau_M$), a well-defined form of dynamic heterogeneity [53] that exists both in the cooled liquid state above $T_g$ and in the glass state well below $T_g$. This puts us in a position to investigate relaxation in the metallic glass and how relaxation is altered by material deformation. In the low temperature non-ergodic glass state. In the glass-forming liquids, it was previously observed that the $\alpha$-structural relaxation time ($\tau_\alpha$) from the long-time decay of the intermediate scattering function $F_s(q, t)$ seems to approach the caging time $t_{cage}$ at high rates of deformation and correspondingly we investigated whether a similar trend might apply for the corresponding relaxation time $\tau_{JG}$ of the deformed metallic glass. We found that $\tau_{JG}$ indeed seems to approach the caging time in the limit of large deformation.

3. The local values of $<u^2>$ have long been interpreted physically as a measure of material local 'stiffness' in relation to the shear modulus of the material and at the scale of chemical bonds when values of $<u^2>$ at an atomic scale are estimated. This suggested the general use of $<u^2>$ as a measure of stiffness at any scale, although this usage of the term means that we may not think of this quantity as directly a measure of shear modulus. We first showed a strong correlation between the macroscopic elastic constant $C_{11}$ as determined from the stress resulting from a low amplitude strain of the metallic glass material and the 'stiffness' measure $k_B T / <u^2>$. We then have firm evidence that $<u^2>$ provides good estimate of the shear stiffness of the material over a large temperature range, encompassing remarkably even the low temperature non-ergodic regime. We also considered estimates of local estimates of stiffness based on $<u^2>$ and local variations of stress and strain and an associated 'local elastic constant' $C_{11,i}$. Again, we found good consistency between the $<u^2>$ stiffness estimate of stiffness and the elasticity motivated definition, but, in this case, we can only claim qualitative a correlation since there is no unique way to define a local shear modulus in materials. By its very definition, $<u^2>$ is a good measure of stiffness at a molecular scale so the use of $<u^2>$ as general measure of stiffness on any length scale seems to be well-supported in our metallic glass system.

4. It has long been thought that SB formation in amorphous solids initiates from relatively 'soft' regions in the material at which large scale non-affine deformations become localized. The test of this hypothesis requires an effective means of identifying 'soft' regions and their evolution as the material is deformed to varying degrees, where the metric of 'softness' must also account for the effect of temperature on local material stiffness. We defined a precise local measure, of 'softness' $\zeta$, defined in terms of material stiffness and find a sharp change in $\zeta$ at a critical strain value at which shear banding initiates. In particular, we find that the critical strain condition for SB formation occurs when the softness ($<u^2>$) distribution within the emerging soft regions approaches that of the interfacial region in its undeformed state, initiating an instability. Correspondingly, no SBs arise when the material is so thin that the entire material can be approximately described as being 'interfacial' in nature.

5. The development of a SB, as in the case of crack formation, involves the emergence of an interface within the material. Correspondingly, we found that the dynamics within the incipient SB region is similar to the free interfacial region of the material, again supporting the existence of an interfacial instability.

6. Based on an observed relationship between the mobile particle cluster lifetime [21] and many observations correlating the JG relaxation process to hypothetical shear transformation zones in glass materials [20], we make the tentative hypothesis that the mobile particle clusters provide well-defined realization of the hypothetical STZs. This identification remains to be tested to validate other attributes of the STZs that have been suggested by recent studies, such as a multipole stress field that plays a large role in the organization of the shear bands. [23,24]

7. We briefly examined the dynamics in the deformation range beyond the onset of shear band formation and find that the structure and dynamics of the material becomes very complex in this region, and we plan to discuss this regime in some detail in a subsequent paper. In the present work, however, we emphasize some significant changes in the nature of the atomic displacement distribution as one undergoes shear band formation. In particular, we find that at large deformations, particle displacement becomes *super-diffusive* beyond a caging regime and the atoms remain localized and nearly Gaussian in a fashion similar to crystalline materials in the caging regime at shorter times. The fully formed shear band regions become highly dynamically heterogeneous (a phenomenon quantified by researchers before us [125,126,163]) and here we focus on the change in the probability of particle displacement, the so called Van Hove function $G_s(r,t)$, as one passes through the SB transition. We find that $G_s(r,t)$ develops an extended 'fat tail' near the yield stain at which shear bands emerge where this extended tail quantifies the greatly enhanced probability for the atoms in the shear band to exhibit jumps to large distances. We interpret this 'fat tail' to arise from the particles being convicted along by the movement of large-scale vortices in the shear band region, as in diverse other fluids exhibiting 'turbulence'. Since the shear band state corresponds to non-equilibrium state similar in some ways to turbulent fluids, we fit $G_s(r,t)$ to a general functional form [98] ($[1 - \beta (1 - q) r^\delta]^{1/(1-q)}$ that commonly arises in probability distributions in turbulent systems and we find that this type of functional form fits our $G_s(r,t)$ very well, but as in turbulent fluids where the same type of fitting has often been made previously, the physical meaning of the fitting parameters is not clear. These observations nonetheless hint at some sort of 'universality' in the atomic dynamics in the shear band state and provide insights into the thermodynamic nature of the non-ergodic glass state. We look forward to further quantifying this novel form of 'elastic turbulence' in the future. Our findings seem to confirm the suggestion made long ago by Taylor [76] that the heterogeneities of turbulent liquids and the formation of heterogeneous structures ("aggregates') in the deformation of polycrystalline materials should have much in common. Our observations also appear complementary to those of Dauchet and Bertin [184] who have recently emphasized the strong analogies between glass-formation and the transition to turbulence in fluids at high Reynolds number based on a combination of phenomenology, energy landscape and dynamical systems ideas.

**Data Availability Statements**

The data that supports the findings of this study are available within the article [and its supplementary material].

**Supplementary Material**

Defines mobile particles, and the spatial and size distribution, fractal geometry of mobile particle clusters in both the interior and interfacial regions of the MG material, protocol to eliminate stress fluctuation, and turbulent-like state of the shear band region.

**Acknowledgements**

H.Z. and X.Y.W. gratefully acknowledge the partial support of the Natural Sciences and Engineering Research Council of Canada under the Discovery Grant Program (RGPIN-2017-03814) and Accelerator Supplements (RGPAS-2017- 507975).

## Conflict of Interest

The authors have no conflict of interests to disclose.

# Supplementary Information:

# The Initiation of Shear Band Formation in Deformed Metallic Glasses from Soft Localized Domains

*Xinyi Wang[1], Hao Zhang[1] †, Jack F. Dougla[2] †*

[1] Department of Chemical and Materials Engineering, University of Alberta, Edmonton, Alberta, Canada, T6G 1H9

[2] Material Measurement Laboratory, Materials Science and Engineering Division, National Institute of Standards and Technology, Gaithersburg, Maryland, USA, 20899

† Corresponding authors: hao.zhang@ualberta.ca; jack.douglas@nist.gov

## A. Definition of Mobile Particles

We follow the methodology developed by Starr et al. [1] to estimate the fraction of 'mobile particles'. In particular, we first bin particles according to the magnitude of their relative displacement as a function of time for a range of threshold fractions, e.g., the 1 % fraction of particles having greatest displacement, the 2 % fraction of particles having greatest displacement, etc. The particles in these classes are 'relatively mobile' to a degree that depends on the cut-off fraction. Clustering of these relatively mobile particles can be defined for each cut-off value as groups of these particles whose nearest-neighbor distance is less than the nearest-neighbor distance defined with reference to the radial distribution function. Following Starr et al., the effect of trivial clusters that exist for randomly located particles is accounted for in the definition of the average cluster mass of these relatively mobile particles. In particular, the average mass of the mobile clusters is normalized by the cluster mass of the same fraction of particles chosen at random. [1] The normalized mass of the relatively mobile particles exhibits a peak for each cut-off value and a *unique* cut-off value can be determined by the particular cut-off size at which the mobile cluster mass is maximized. If the cut-off had been chosen too small then one only determines fragments of the mobile particles of interest and if the cut-off is chosen too large, then one starts to incorporate particles that exist because of clustering at random. The methodology allows one to zero in on the average cut-off value in a systematic and objective fashion. The data in Fig. S.1 indicates the normalized maximum cluster size maximizes at a cut-off value of 2.75 % for this material. Specifically, we define 'mobile particles' as those particles that have the top 2.75 % displacement at any time $t$. The time at which the average mass of the mobile particle peaks is the mobile particle lifetime, $\tau_M$. See Starr et al. for illustration of the time dependence of the mobile particle mass.[1] The near linear relation between $\tau_M$ and the peak in the non-Gaussian parameter $t^*$ shown in Fig.

5(a) of the main paper relies only on holds when the fraction of mobile particles is estimated with some precision (roughly about 2.75 % ± 1.0 %) because the mobile particle lifetime $\tau_M$ slightly depends on the choice of cut-off for defining mobile particles.

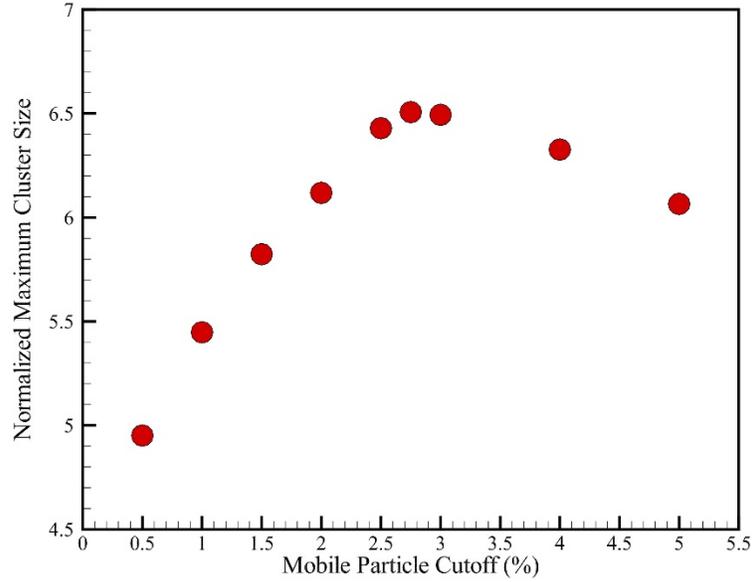

**Figure S1.** Normalized maximum cluster size as a function of mobile particle cutoff when $\varepsilon = 7.0$ % in a sample with $h = 300$ Å at $T = 50$ K.

## B. Spatial and size distribution and fractal geometry of mobile particles in the interior and interfacial regions

We next examine the geometry of the mobile particles in interior and interfacial regions. We calculate the fractal dimension, $d_f$, to characterize the geometrical properties of mobile clusters in both regions. The 'fractal' dimension of the mobile clusters is calculated using the relation, $n \sim R_g^{d_f}$, where $n$ is the number of particle clusters and $R_g$ is the radius of gyration of the given clusters which is calculated using the equation: $R_g^2 = \frac{1}{2N}\sum_{i,j}(r_i - r_j)^2$ ($N$ is the total number of mobile particles in the cluster and $r_i$, $r_j$ are the positions of the $i^{th}$ and $j^{th}$ particles). As shown in the inset of Fig. S2, we found out that $d_f$ varies between 2.6 and 2.9 in the strain range and regions that we studied, similar to the findings in our previous studies on mobile clusters in the metallic glasses systems at temperatures above the glass transition temperature, $T_g$ [2]. Next, we consider the size distribution of the mobile particle clusters in both regions. $P(n)$ can be described using a power law $P(n) \sim n^{-\tau_F}$ where $\tau_F$ is the size distribution scaling exponent. As illustrated in Fig. S3, we see that $\tau_F$ around 1.5 in both regions. The size distribution exponent ('Fisher exponent') for mobile particle clusters is also nearly the same as we have seen before [2] for the same metallic glass system for a $T$ range much greater than $T_g$ so there is apparently no essential difference in the geometry of the mobile particle clusters above and below $T_g$.

We show some randomly selected mobile particle clusters in the interior and interfacial regions of the metallic glass (MG) in the inset where it is apparent that these clusters take a relatively compact and symmetric form. This is natural since a perfectly compact object, like a sphere, has a fractal dimension $d_f = 3$. Notice that the average mass of these clusters ranges from about 10 to 500. Based on our discussion in the main text indicating that the lifetime of the mobile particles can be identified with the Johari-Goldstein (JG) relaxation time and the commonly stated correlation between the JG relaxation process and shear transformation zones (STZs), we suggest that the 'mobile particle clusters' can be identified as being a concrete realization of STZs [3–5]. Recent atomic resolution imaging of the interface of metallic glass and other glass-forming liquids using ultrahigh vacuum scanning microscopy have indicated the presence of relatively compact mobile particle clusters on the surface of glasses deep in their glass state where the dynamics of their rearrangement motion was suggested to be consistent with the JG relaxation process and where the size of the clusters was estimated to be in the range of 4 to 5 atom diameters [6–8]. The appearance and reported size of these clusters seem to be consistent with our simulations. Further comparisons between simulation and these clusters observations seem warranted.

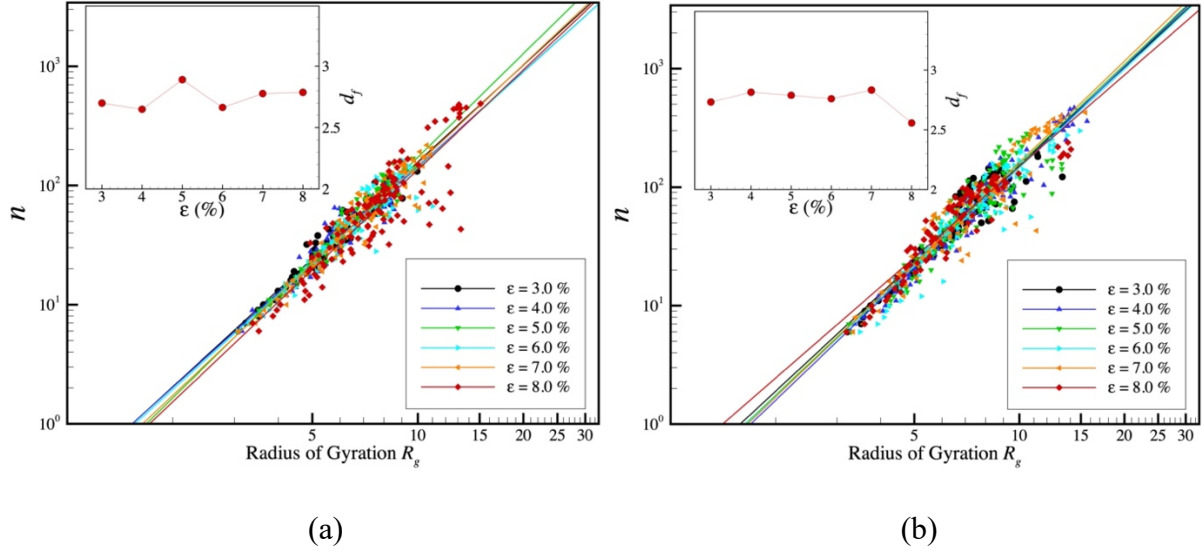

(a)                                          (b)

**Figure S2.** Cluster radius of gyration $R_g$ with its mass $n$, $n \sim R_g^{d_f}$ at different strains in the interior (a) and interfacial (b) regions. The inset shows that the fractal dimensions, $d_f$, do not vary significantly with strain or regions. The fractal dimensions in both regions are around 2.8.

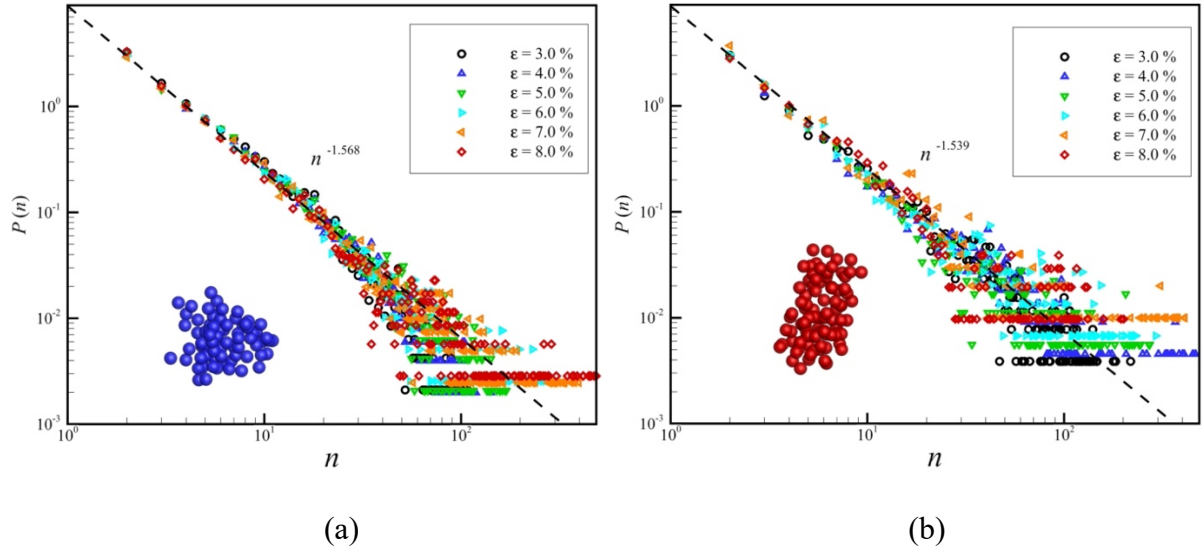

(a)                                          (b)

**Figure S3.** The distribution of mobile particle cluster sizes $P(n)$ in the interior (a) and interfacial (b) regions at different strains. The distribution can be described using a power law with size distribution exponent (Fisher exponent) $\tau_F$ around 1.5. Atomic configurations of representative clusters in both regions at $\varepsilon$ = 8.0 % are also shown.

## C. New Protocol to Eliminate Stress Fluctuation

As shown in Figure 1(a) in the main text, the stress fluctuations in the large deformation regime are well-known and expected for glass materials beyond their point of "yield" and this is naturally attributed to the highly heterogeneous nature of the material in its shear banded state. The fluctuations at low strain are another matter. We did not expect "fluctuation effects" in this regime. The magnitude of the estimated $<u^2>$ showed some fluctuations in this low deformation regime so we considered further what these "fluctuations" at low $\varepsilon$ might represent.

It is likely that the quench process of our metallic glass might lead to appreciable residual stresses in the low temperature that might be relieved small amplitude deformations. There have been many experimental and computational reports recently of how the application of small stresses can alter the "energy landscape" of glassy materials so this seemed like a plausible reason for the appreciable stress fluctuation effects that we observe at very low deformations. To check this hypothesis, we subjected the material to a very small prestress (deformation) to allow these hypothetical residual stresses to relax and then took the resulting material to be our relaxed undeformed material. This procedure not only essentially eliminated the stress fluctuations at low $\varepsilon$ (see inset to Fig. S4), this pre-stress procedure also essentially eliminated the "noisy" nature of the average value of $<u^2>$ data at low deformation. The average value of $<u^2>$ at $\varepsilon = 0$ % is now quite consistent with the average value of predicted from expression in the paper.

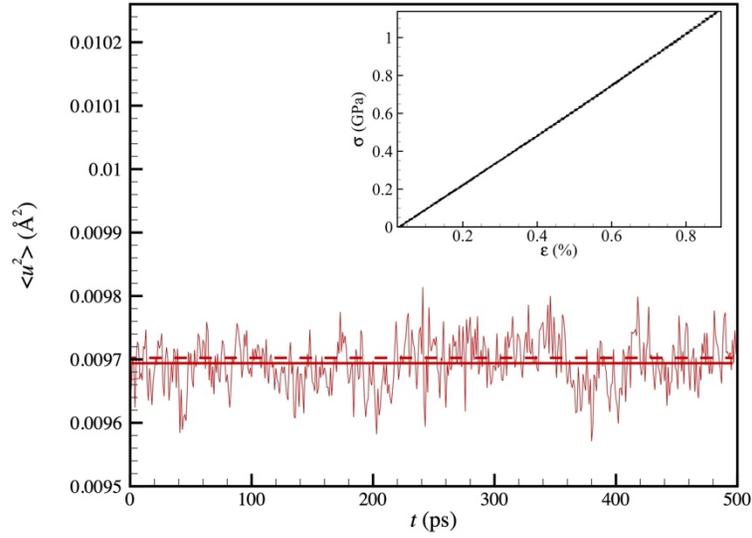

**Figure S4.** Stress versus strain in deformed ZrCu material subjected to pre-stress and resulting time series of $<u^2>$ as a function of time $t$ and $\varepsilon = 0$. The power spectrum of $<u^2>$ fluctuation in the undeformed state exhibits a power-law scaling with frequency, where the color noise exponent is equal to 0.35. The dashed horizontal line represents the predicted average value of $<u^2>$ from the expression for $<u^2>$ versus $\varepsilon$ given our paper and the solid horizontal line represents the average value of $<u^2>$ over 500 ps in the undeformed state. The inset of this paper shows the stress versus strain relation of the pre-stressed material where we see that the "fluctuations" observed before have now been essentially eliminated.

## D. Turbulent-like State of the Shear Band Region

Figure S5 shows the local shear modulus distribution captured from the system with $h = 300$ Å which contains part of the SB shown in blue, with the particle displacement vector in the corresponding region.

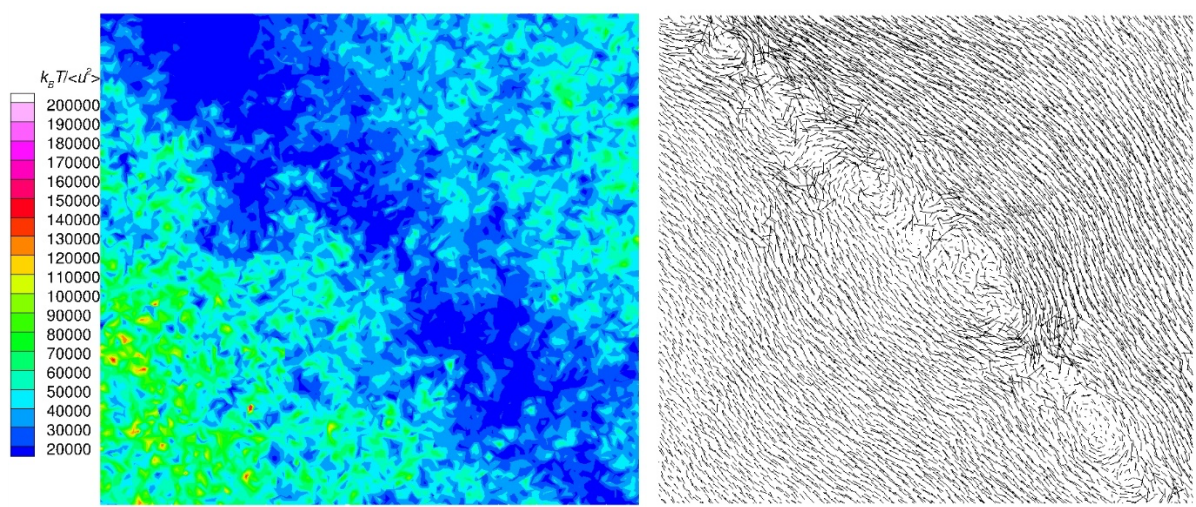

**Figure S5.** Contour map of local shear modulus $k_B T / <u^2>$ around shear band in the system with thickness $h = 300$ Å, and displacement vector in the corresponding region at $\varepsilon = 8.9$ %, corresponding to shear banded state. From the corresponding particle displacement vector, a circular collective, and vortex-like motion has been observed, similar to the findings from Sopu's work [9]. This is also a common type of collective motion seen in dense granular materials [10]. This type of vortex-like motion has been previously reported in the SB region and these coherent vortex-like structures were interpreted to arise from the percolation of smaller STZs structures [9]. Note the vortex patterns involving collective particle displacement that are also characteristic of turbulent fluids and other materials that have been characterized by analogy as being "turbulent". We plan to quantify this vortex-like motion in a separate publication using methods utilized for other "turbulent" driven materials. Figure S5 is intended to only qualitatively illustrate the complex particle displacements in the shear banded material.